\documentclass[11pt,a4paper]{article}
\usepackage[UKenglish]{babel}
\usepackage[UKenglish]{isodate}
\usepackage[T1]{fontenc}
\usepackage{lmodern}
\usepackage[top=0.75in,left=0.75in,right=0.75in,bottom=1.0in]{geometry}
\cleanlookdateon
\usepackage{graphicx}
\usepackage[margin=15pt,font={footnotesize},labelfont=bf,format=hang]{caption}
\usepackage[margin=15pt,font={footnotesize},labelfont=rm,format=hang,labelformat=simple]{subcaption}

\usepackage{bm}
\usepackage{amsbsy}
\usepackage{amssymb}
\usepackage{amsmath}
\usepackage{mathtools}
\usepackage{isomath}
\usepackage{enumitem}
\usepackage{natbib}
\usepackage[pdfencoding=auto]{hyperref}

\usepackage[dvipsnames]{xcolor}
\usepackage{cleveref}
\crefname{equation}{}{}
\Crefname{equation}{Equation}{Equations}
\crefname{section}{\S}{\S\S}
\Crefname{section}{Section}{Sections}
\usepackage{booktabs}
\usepackage{siunitx}
\usepackage{authblk}
\usepackage{microtype}
\usepackage{tikz}

\renewcommand{\v}[1]{\mathbfit{#1}}      
\newcommand{\uv}[1]{\widehat{\mathbfit{#1}}}      
\newcommand{\q}[1]{\mathsfbfit{#1}}        
\renewcommand{\t}[1]{\mathsfbfit{#1}}	 
\newcommand{\vu}[1]{\mathbf{#1}}         
\newcommand{\tu}[1]{\bm{\mathsf{#1}}}	 

\newcommand{\upi}{\pi}

\newcommand{\qprod}{\mathbin{\bullet}} 

\renewcommand{\d}{{\mathrm d}}
\def \p {\partial}

\newcommand{\nx}{t}
\newcommand{\ny}{\mu}
\newcommand{\nz}{\nu}

\renewcommand{\tilde}{\widetilde}

\newcommand{\fd}[2]{\mathchoice{\frac{\d #1}{\d #2}}{\d #1/\d #2}{\d #1/\d #2}{\d #1/\d #2}}
\newcommand{\pd}[2]{\mathchoice{\frac{\p #1}{\p #2}}{\p #1/\p #2}{\p #1/\p #2}{\p #1/\p #2}}

\DeclareMathSymbol{\Delta}{\mathalpha}{operators}{1} 
\newcommand{\Sp}{\text{\rm \textit{Sp}}}

\newcommand{\trans}{\mathsf{T}}

\begin{document}

\title{\bf Undulatory swimming in suspensions and networks of flexible filaments}
\author[1]{Adam K Townsend\thanks{Corresponding author: \url{adam.k.townsend@durham.ac.uk}}}
\author[2]{Eric E Keaveny}
\affil[1]{\it\normalsize Department of Mathematical Sciences, Durham University, Stockton Road, Durham DH1 3LE, UK}
\affil[2]{\it\normalsize Department of Mathematics, Imperial College London, London SW7 2AZ, UK}
\maketitle

\begin{abstract}
Many biological fluids are composed of suspended polymers immersed in a viscous fluid.  A prime example is mucus, where the polymers are also known to form a network.  While the presence of this microstructure is linked with an overall non-Newtonian response of the fluid, swimming cells and microorganisms similar in size to the network pores and polymer filaments instead experience the heterogeneous nature of the environment, interacting directly with the polymers as obstacles as they swim.  To characterise and understand locomotion in these heterogeneous environments, we simulate the motion of an undulatory swimmer through 3D suspensions and networks of elastic filaments, exploring the effects of filament and link compliance and filament concentration up to 20\% volume fraction.  For compliant environments, the swimming speed increases with filament concentration to values about 10\% higher than in a viscous fluid.  In stiffer environments, a non-monotonic dependence is observed, with an initial increase in speed to values 5\% greater than in a viscous fluid, followed by a dramatic reduction to speeds just a fraction of its value in a viscous fluid.  Velocity fluctuations are also more pronounced in stiffer environments.  We demonstrate that speed enhancements are linked to hydrodynamic interactions with the microstructure, while reductions are due to the filaments restricting the amplitude of the swimmer's propulsive wave.  Unlike previous studies where interactions with obstacles allowed for significant enhancements in swimming speeds, the modest enhancements seen here are more comparable to those given by models where the environment is treated as a continuous viscoelastic fluid.\end{abstract}

\tableofcontents

\section{Introduction}

The fluids produced and used by living organisms are inherently complex, containing a suspended microstructure such as deformable cells or polymers molecules that are immersed in a viscous fluid.  Prime examples of such fluids are blood and mucus.  While blood can be viewed as a suspension of deformable cells, mucus is instead a mesh of fibre bundles formed from the protein mucin and connected through entanglement or hydrophobicity \citep{lai_altering_2009}.  In our own bodies, mucus plays several important roles as a lubricating fluid \citep{boegh_mucus_2015,nordgard_oligosaccharides_2011}, a protective layer \citep{allen_gastroduodenal_2005,mantelli_functions_2008}, and as a barrier to pathogens and infection \citep{ensign_nanoparticle-based_2014,johansson_inner_2008}.  Mucus helps facilitate digestion, protecting the soft tissue of the stomach from the acids used during digestion, and from bacteria such as {\it H. pylori} that cause stomach ulcers \citep{celli_helicobacter_2009}.  In relation to reproduction, mucus is present in the cervix \citep{ceric_ultrastructure_2005,rutllant_ultrastructural_2005}, again serving to prevent infections, but also providing the fluid environment through which sperm swim.  Mucin fibre density is known to decrease when fertilisation is possible, and sperm--mucus interactions have been put forward as a possible mechanism for selection of viable sperm \citep{bianchi_human_2004,suarez_sperm_2006,holt_sperm_2015}.

As one might expect, the presence of the microstructure, which produces changes in the rheological properties of the surrounding fluid, alters the motility of cells moving through the fluid \citep{sznitman_locomotion_2015}.  Sperm cells swim by undulating their flagellum, and in purely viscous fluids, increased viscosity is seen to be associated with lower wave propagation velocity and smaller wave amplitude in sperm, but not necessarily with any change in average swimming speed \citep{brokaw_effects_1966,smith_bend_2009}. But human cervical mucus is viscoelastic and shear-thinning \citep{wolf_human_1977}, and increased viscosity in mucus \emph{is} linked to enhanced swimming speed \citep{lopez-gatius_sperm_1994,rutllant_ultrastructural_2005}. Conversely, experiments with \emph{C.\ elegans}, which also moves through undulation, show that increased fluid elasticity leads to slower swimming speeds \citep{shen_undulatory_2011}. As increasing the microstructure in the fluid changes the fluid viscoelasticity, in general, the relationship between suspension concentration and microorganism motility is not simple. As another example, experiments show that \emph{E.\ coli}, which propels itself through rotation of its helical flagella, travels faster in low-concentration polymeric fluids (up to about 20\%) than in a purely viscous fluid, but at high concentrations of polymers, travels slower \citep{schneider_effect_1974,martinez_flagellated_2014}.

To gain a more fundamental understanding of microorganism motility in complex biological fluids, researchers have turned to mathematical models based on continuum-level descriptions of non-Newtonian fluids.  Initial theoretical work with Oldroyd-B and FENE-P viscoelastic models, on infinitely long undulating sheets and filaments, typically predicted that additional viscoelasticity would impede swimming \citep{lauga_propulsion_2007,fu_swimming_2009}. 
Later, computational studies in the nonlinear regime for finite-length undulatory swimmers painted a more complex picture, with with both hindrance and enhancement (up to 30\%) found for different mechanical properties of the swimmer in a two-dimensional Oldroyd-B fluid \citep{teran_viscoelastic_2010,thomases_mechanisms_2014}. 
The same broad picture was found in 2D asymptotic and numerical work modelling the background fluid with the Brinkman equations for porous media \citep{leiderman_swimming_2016}, with both hindrance and enhancement (up to 45\%) found for finite-length swimmers with low and high bending stiffness respectively. The same model in three dimensions also showed hindrance and enhancement (up to 15\%) \citep{cortez_computation_2010,olson_effect_2015}, depending on bending stiffness.

While these models provided a number of key insights into how fluid rheology affects microorganism motility, the continuum fluid model assumes a large separation in length scale between the swimming microorganism and the microstructure, describing just one limit in a much larger parameter space (`type II' in \citet{spagnolie_swimming_2023}).  In many notable cases, the characteristic length scales of the swimming body and the immersed microstructure are comparable (the border between `types II and IV').  For sperm moving in cervical mucus, for example, the gap sizes in the mucin fibre bundle mesh ($\sim$\SI{1}{\micro\metre}) are on the same order as the head of the cell ($\sim$\SI{5}{\micro\metre})  \citep{sheehan_electron_1986,lai_altering_2009}.  In this scenario, it is therefore important to consider the fluid microstructure as discrete objects comparable in size to the swimmer.  
One intermediate method is to represent these discrete objects as point forces which act on the swimmer, where the spacing between the forces (e.g.\ in a lattice or a mesh) is of the same order as the swimmer size \citep{gniewek_coarse-grained_2010}. Simulations of finite-length swimmers through such a mesh network have found mostly hindrance \citep{schuech_performance_2022} but also speed enhancement of up to 30\% \citep{wrobel_enhanced_2016}.  Simulations of infinitely long swimmers through bead--spring dumbbell suspensions \citep{zhang_reduced_2018} have shown both hindrance and enhancement (up to 10\%) for sufficiently short-wavelength swimmers, depending on the relative viscosity of the fluid around the cell body.  
Experiments and simulations \citep{park_enhanced_2008,majmudar_experiments_2012} of undulatory locomotion in lattices of rigid obstacles (`type I') have shown that speeds of finite-length swimmers can increase by nearly a factor of ten when the lattice spacing allows for the swimmer to constantly push itself along as it undulates.  After introducing randomness in the obstacle positions and environment compliance through tethering springs \citep{kamal_enhanced_2018}, enhanced locomotion continues to be observed, though by a more modest but still substantial factor of two and a half.  In addition, modelling the microstructure discretely also captures swimmer velocity fluctuations \citep{jabbarzadeh_swimming_2014,kamal_enhanced_2018} due to collisions which in turn produces effective swimmer diffusion over long times.  Non-trivial changes in motility are also observed for swimmers propelled by helical flagella in heterogeneous environments.  A recent theory to explain increased swimming speeds at low suspension concentrations is that the swimming motion of the microorganism leads to phase separation near the surface of the swimmer. Multiparticle collision dynamics simulations of finite-length helically propelled swimmers \citep{zottl_enhanced_2019} suggest this could lead to speed increases between 10\% and 100\%.  This mechanism yields similar increases in swimming speed for infinitely long undulatory swimmers \citep{man_phase-separation_2015}.  For recent reviews of theoretical and computational approaches to modelling swimming in complex media, see \citet{li_microswimming_2021,spagnolie_swimming_2023}; for a broad overview of sperm modelling, see \citet{gaffney_modelling_2021}.

In this paper, we perform simulations of an undulatory swimmer moving through three-dimensional suspensions of elastic filaments.  We resolve hydrodynamic interactions using the force coupling method and elastic deformation of the swimmer and filaments using the computational framework described in \citet{schoeller_methods_2021}.  A summary of the model is presented in \cref{sec:model}.  In performing the simulations, we consider cases where the filaments are freely suspended in the fluid (\cref{sec:soups}), or tethered to one another via Hookean springs to form a network (\cref{sec:networks}).  In both types of environments, we examine how the swimming speed is affected by filament concentration, filament stiffness, and tether stiffness.  In sufficiently compliant environments, we find that the swimming speed increases to a value of 10\% greater than that in viscous fluid for volume fractions up to 20\%.  For stiff environments, however, the speed in non-monotonic with volume fraction, with peak enhancement of 5\% and reductions in speed as the volume fraction approaches 20\%.  By systematically removing physical effects from the simulation, we show that speed enhancement is due to the hydrodynamic interactions, where as speed reductions in stiff environments are due to reduced waveform amplitude as the swimmer becomes more confined.  

\section{Mathematical model for the swimmer and environment}
\label{sec:model}

\subsection{General model for all filaments}

In our study, we simulate the movement of an active, undulating swimmer, modelled as an active filament, travelling through a fully three-dimensional suspension or network of passive filaments.  To perform these simulations, we employ the filament model developed in \citet{schoeller_methods_2021}. The model is a three-dimensional development of the two-dimensional model of swimming filaments first introduced in \citet{majmudar_experiments_2012}, adapted for sperm in \citet{schoeller_flagellar_2018}, and used in studies of two-dimensional obstacle arrays in \citet{kamal_enhanced_2018}.  For full discussion, we refer the reader to \citet[\S\,2]{schoeller_methods_2021}; we will summarise the model here in the context of a single filament.  
	
\begin{figure}
    \centering
    \begin{tikzpicture}
        \node[anchor=north west,inner sep=0] (image) at (0,0) {\includegraphics[scale=0.92]{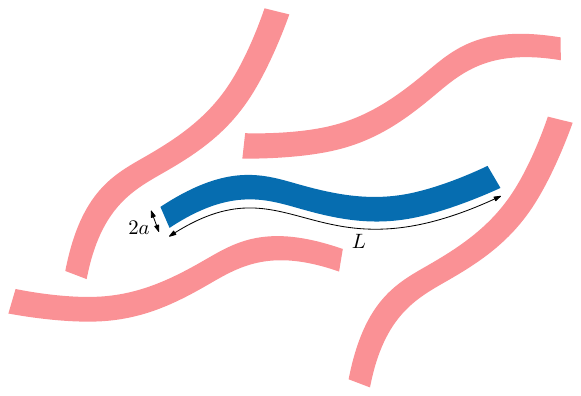}};
        \node[anchor=north west,inner sep=0] at (0,0) {(\textit{a})};
    \end{tikzpicture}
    \vspace{3mm}

    \begin{tikzpicture}
        \node[anchor=north west,inner sep=0] (image) at (0,0) {\includegraphics[scale=0.92]{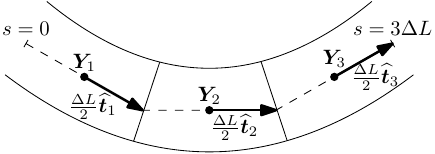}};
        \node[anchor=south west,inner sep=0] at (0,0) {(\textit{b})};
    \end{tikzpicture}
    \quad
    \begin{tikzpicture}
        \node[anchor=north west,inner sep=0] (image) at (0,0) {\includegraphics[scale=0.92]{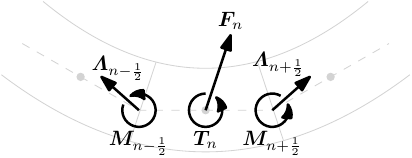}};
        \node[anchor=south west,inner sep=0] at (0,0) {(\textit{c})};
    \end{tikzpicture}    
    \caption{Filament context and discretisation. (\textit{a}) Cartoon of the swimming suspension: the swimmer (blue) makes its way through a suspension of passive network filaments (pale red) in 3D. (\textit{b}) Spatial discretisation of the filament with positions $\v{Y}_n$ and vectors $\uv{t}_n$ satisfying the constraint, \cref{tangent-vector}. (\textit{c}) Forces and torques in the discrete system with $\v{F}_n = \Delta L \v{f}_n$ and $\v{T}_n = \Delta L \v{\tau}_n$. Diagrams reproduced from \citet{schoeller_methods_2021}.}
    \label{fig:cartoon}
\end{figure}
	
In the model, the filament, sketched in \cref{fig:cartoon}, has length $L$ and a circular cross-section of radius $a$.  The filament is parametrised by its arclength, $0 \leq s \leq L$, such that the position of any point on the filament centreline at time $t$ is $\v{Y}(s,t)$. Additionally, we introduce a right-handed orthonormal frame at any $s$ and $t$, $\{\uv{t}(s,t),\uv{\mu}(s,t),\uv{\nu}(s,t)\}$, that we will employ to keep track of the deformation of the filament. 

To derive the equations of motion, we begin by considering the force and moment balances along the filament 
\citetext{\citealp[\S\,19]{landau_theory_1986}; \citealp{powers_dynamics_2010}},
\begin{align}
    \pd{\v{\Lambda}}{s} + \v{f} &= \vu{0}, \label{eq:KirchhoffForce}\\
    \pd{\v{M}}{s} + \uv{t} \times \v{\Lambda} + \v{\tau} &= \vu{0},\label{eq:KirchhoffMoment}
\end{align}
where $\v{\Lambda}(s,t)$ and $\v{M}(s,t)$ are the internal force and moment on the filament cross-section.  The internal moments, $\v{M}(s,t)$, are linearly related to the filament bending and twist through the constitutive law
\begin{align} 
\label{equation:3d_moment_frame}
  	\v{M}(s,t)= K_B \left[\uv{\ny}\left(\uv{t}\cdot\pd{\uv{\nz}}{s} - \kappa_\ny(s,t) \right) + \uv{\nz}\biggl(\uv{\ny}\cdot\pd{\uv{t}}{s}- \kappa_\nz(s,t)\biggr)\right] + K_T \uv{t} \left(\uv{\nz}\cdot \pd{\uv{\ny}}{s} -\gamma_0(s,t) \right),
\end{align}
where $K_B$ is the bending modulus and $K_T$ is the twist modulus.  We also have preferred curvatures $\kappa_{\ny}(s,t)$ and $\kappa_{\nz}(s,t)$ and preferred twist, $\gamma_0(s,t)$, that can incorporate nontrivial equilibrium filament shapes \citep{lim_dynamics_2008, lim_dynamics_2010, olson_modeling_2013} and time-dependent filament motion.  For the passive filaments comprising the suspension or network, the preferred curvatures and twist are zero, but for the swimmer, there is a time-dependent $\kappa_\nz$, as defined in \cref{sec:simulation-parameters}.  The internal force, $\v{\Lambda}(s,t)$, enforces the kinematic constraint
\begin{equation}
	\uv{t} = \pd{\v{Y}}{s}; \label{tangent-vector}
\end{equation}
unlike the internal moment, the internal force is \emph{a priori} unknown and will be solved for in the final part of the numerical method.

Also appearing in \cref{eq:KirchhoffMoment} are $\v{f}$ and $\v{\tau}$ which are the external force and torque per unit length on the filament, respectively.  These consist of short-ranged, repulsive forces when filaments collide, but also the force and torque per unit length exerted on the filament by the surrounding fluid.
		    
Next, we discretise the filament into $N$ segments of length $\Delta L$. Each segment has position $\v{Y}_n$ and orientation vector $\uv{t}_n$. After applying central differencing to \cref{eq:KirchhoffMoment}, we have
\begin{align}
    	\frac{\v{\Lambda}_{n+1/2} - \v{\Lambda}_{n-1/2}}{\Delta L} + \v{f}_n &= \vu{0}, \label{eq:DifferencingForce} \\
    	\frac{\v{M}_{n+1/2} - \v{M}_{n-1/2}}{\Delta L} + \frac{1}{2} \uv{t}_n \times \left(\v{\Lambda}_{n+1/2} + \v{\Lambda}_{n-1/2}\right) + \v{\tau}_n &= \vu{0}, \label{eq:DifferencingMoments}
\end{align}
where $\v{M}_{n+1/2}$ and $\v{\Lambda}_{n+1/2}$ are the internal moment and force between segments $n$ and $n+1$, and $\v{f}_n$ and $\v{\tau}_n$ are the external force and torque, respectively, per unit length on segment $n$.  For the discrete system, $\v{\Lambda}_{n+1/2}$ can be viewed as the Lagrange multiplier that enforces 
\begin{equation}
	\v{Y}_{n+1} - \v{Y}_n - \frac{\Delta L}{2}\left( \uv{t}_n + \uv{t}_{n+1} \right) = \vu{0},
	\label{lagrange-multiplier-equation}
\end{equation}
the discrete version of \cref{tangent-vector}.

Multiplying \cref{eq:DifferencingForce,eq:DifferencingMoments} through by $\Delta L$, we arrive at the force and torque balances on each segment $n$,
\begin{align}
	\v{F}^C_n + \v{F}_n &= \vu{0}, \label{eq:DiscreteForce} \\
    \v{T}^E_n + \v{T}^C_n + \v{T}_n &= \vu{0}, \label{eq:DiscreteTorque}
\end{align}
where $\v{F}^C_n =\v{\Lambda}_{n+1/2} - \v{\Lambda}_{n-1/2}$, $\v{T}^E_n = \v{M}_{n+1/2} - \v{M}_{n-1/2}$, and $\v{T}^C_n = \frac{\Delta L}{2} \uv{t}_n\times(\v{\Lambda}_{n+1/2} + \v{\Lambda}_{n-1/2}).$	The external forces and torques acting on each segment are $\v{F}_n = \Delta L \v{f}_n$ and $\v{T}_n = \Delta L \v{\tau}_n$ and are composed of the repulsive, or barrier, forces between segments and the hydrodynamic force and torque on each segment, $\v{F}_n = \v{F}^B_n - \v{F}^H_n$ and $\v{T}_n = - \v{T}^H_n$.  The barrier force, $\v{F}^B_n$, prevents segments from overlapping and acts between pairs of segments of different filaments, or between non-neighbouring segments of the same filament.  Specifically, expressing the centre-to-centre displacement between two segments as $\v{r}_{nm} = \v{Y}_n - \v{Y}_m$, then $\v{F}^B_n = \sum_m \v{F}^B_{nm}$ where \citep{dance_collision_2004} 
\begin{align}
    \v{F}^B_{nm} = F^S \left(\frac{4a^2\chi^2 - r_{nm}^2}{4a^2(\chi^2 - 1)}\right)^4 \frac{\v{r}_{nm}}{2a} \quad  (\text{if } r_{nm} < 2 \chi a),
    \label{eq:barrier-force}
\end{align}
and $\v{F}^B_{nm} = \vu{0}$ otherwise, where $r_{nm} = |\v{r}_{nm}|$ and $\chi = 1.1$. The force strength $F^S$ is set to $51K_B/L^2$, where $K_B$ and $L$ are the bending modulus and length of the swimmer described in \cref{sec:simulation-parameters}.  The steric forces serve to prevent the filament and swimmer segments from overlapping.  The choice of $F^S = 51K_B/L^2$ is chosen to ensure that overlapping does not occur while also not introducing numerical stiffness that would lead to large increases in the number of Broyden iterations needed during a timestep.  Additionally, this force captures steric interactions between the swimmer and filaments that are expected to occur in real swimmer--biopolymer suspensions where lubrication interactions break down due to polymer roughness and surface aberrations.
    
At this point we now have full expressions for the hydrodynamic forces and torques, $\v{F}^H_n$ and $\v{T}^H_n$, acting on all filament segments, in terms of the unknown Lagrange multipliers $\v{\Lambda}_n$. As such, we now consider the hydrodynamic model which will allow us to close the system and obtain the motion of the filaments using an iterative method.  Due to the negligible influence of fluid inertia, the hydrodynamic interactions are governed by the Stokes equations for a fluid with viscosity $\eta$. Due to the linearity of the Stokes equations, the velocities, $\v{V}_n$, and angular velocities, $\v{\Omega}_n$, of the segments are linearly related to the hydrodynamic forces and torques on the segments, leading to the mobility problem
\begin{equation}\label{eq:mobility-problem}
	\begin{pmatrix}
	\v{V}\\\v{\Omega}
	\end{pmatrix}
    = \mathcal{M}\cdot
    \begin{pmatrix}
	\v{F}^H\\\v{T}^H
	\end{pmatrix},
\end{equation}
where $\v{V}^\trans = (\v{V}_1^\trans,\dots,\v{V}_N^\trans)$ is the vector of all segment velocity components, and similarly for $\v{\Omega}$, $\v{F}^H$ and $\v{T}^H$.  

The computation of the mobility matrix, $\mathcal{M}$, (or, more efficiently, the computation of its action on the force and torque vector) can be performed with a number of different hydrodynamic models and computational methodologies. In this paper, we use the matrix-free Force Coupling Method (FCM) \citep{maxey_localized_2001,lomholt_force-coupling_2003,liu_force-coupling_2009}.  In FCM, the forces and torques exerted on the fluid by the segments are transferred through a truncated, regularised force multipole expansion in the Stokes equations. The delta functions in the multipole expansion are replaced by smooth Gaussians, and, using the ratios established for spherical particles, the envelope size of the Gaussians is based on the filament radius, $a$.
After finding the fluid velocity, the same Gaussian functions are used to average over the fluid volume to obtain the velocity, $\v{V}_n$, and angular velocity, $\v{\Omega}_n$, of each segment, $n$.
	
Having obtained the velocities and angular velocities of the segments, we now integrate in time. For the segment positions, we integrate
\begin{equation}
	\fd{\v{Y}_n}{t} = \v{V}_n,
	\label{position-evolution}
\end{equation}
while for segment orientations, we have 
\begin{equation}
	\fd{\q{q}_n}{t} = \frac{1}{2}(0, \v{\Omega}_n) \qprod \q{q}_n,
	\label{quaternion-evolution}
\end{equation}
where $\q{q}_n$ is the unit quaternion that provides the rotation matrix,
\begin{equation}\label{eq:RotToFrame}
    \t{R}\left(\q{q}_n(s,t)\right) = \left(\uv{\nx}_n\;\; \uv{\ny}_n\;\; \uv{\nz}_n \right).
\end{equation}
To advance in time, we discretise \cref{position-evolution,quaternion-evolution} using a geometric second-order backward differentiation scheme \citep{ascher_computer_1998,faltinsen_multistep_2001,iserles_lie-group_2000} that ensures that the quaternions remain unit length.  The resulting equations, along with \cref{lagrange-multiplier-equation}, yield a system of nonlinear equations that we solve using Broyden's method \citep{broyden_class_1965}, a quasi-Newton iterative method, to obtain the updated segment positions, $\v{Y}_n$, the quaternions associated with the segment orientations, $\q{q}_n$, and the Lagrange multipliers, $\v{\Lambda}_n$. 

For full details of the filament model, especially in regard to the quaternion representation and geometric integration scheme, we again refer the reader to \citet[\S\,2]{schoeller_methods_2021}.

\subsection{Swimmer}
\label{sec:simulation-parameters}

In all our simulations, the swimmer has length $L=33a$, is discretised into $N=15$ segments, and its bending modulus and twisting modulus are equal, $K_T = K_B$.  Swimmer motion is generated by planar undulations, driven by a preferred curvature, 
\begin{equation}
	\kappa_\nz(s,t) = K_0 \sin (ks - \omega t + \phi)\cdot\begin{cases}2(L-s)/L&s>L/2,\\1&s\leq L/2,\end{cases}
\end{equation}
where $k = 2\upi/L$. To connect with previous work in \citet{kamal_enhanced_2018}, we choose $K_0 = 8.25/L$ and set $\omega$ to yield a dimensionless sperm number $\Sp = (4\upi\omega\eta/K_B)^{1/4}L = 5.87$.  The linear decay in amplitude for $s > L/2$ is introduced to limit otherwise very high curvatures toward the end of the swimmer \citep{majmudar_experiments_2012,schoeller_flagellar_2018,kamal_enhanced_2018}.  The other preferred curvature, $\kappa_\ny$, and the preferred twist, $\gamma_0$, are set to zero.  
Under these conditions, in a quiescent fluid, a single swimmer swims its length $L$ in a time of $11.5T$, where $T = 2\upi/\omega$ is the undulation period. The resultant periodic waveform has a wavelength equal to the length of the swimmer, broadly matching observations of sperm cells and \emph{C.\ elegans} \citep{rikmenspoel_tail_1965,sznitman_material_2010}; it is depicted in the left-hand panel of \cref{fig:swimmer_snapshots_overlay_rft}.

\subsection{Domain size}
All simulations are performed in a cubic periodic domain of size $87a\times87a\times87a$ = $2.64L \times 2.64L \times 2.64L$, where $L$ is the swimmer length, with a grid size of $288\times288\times288$ used to solve the Stokes equations as part of the FCM mobility problem. The domain size is chosen to ensure periodic images have a negligible effect on the swimmer's motion. Illustrations of the domain will be presented in the next section.

\section{Unconnected filament suspensions}
\label{sec:soups}

In our simulations, we will consider the motion of the swimmer through two types of filament environments: a suspension of unconnected filaments, and a spring-connected filament network.  In this section, we look at the former, first summarising the approach that we use to generate this environment and then presenting the results of our simulations.

\subsection{Filament environment}\label{sec:filament-environments}
\label{sec:soup-seeding}

\begin{figure}
    \centering
    \begin{tikzpicture}
        \node[anchor=north west,inner sep=0] (image) at (0,0) {\includegraphics[width=0.32\textwidth]{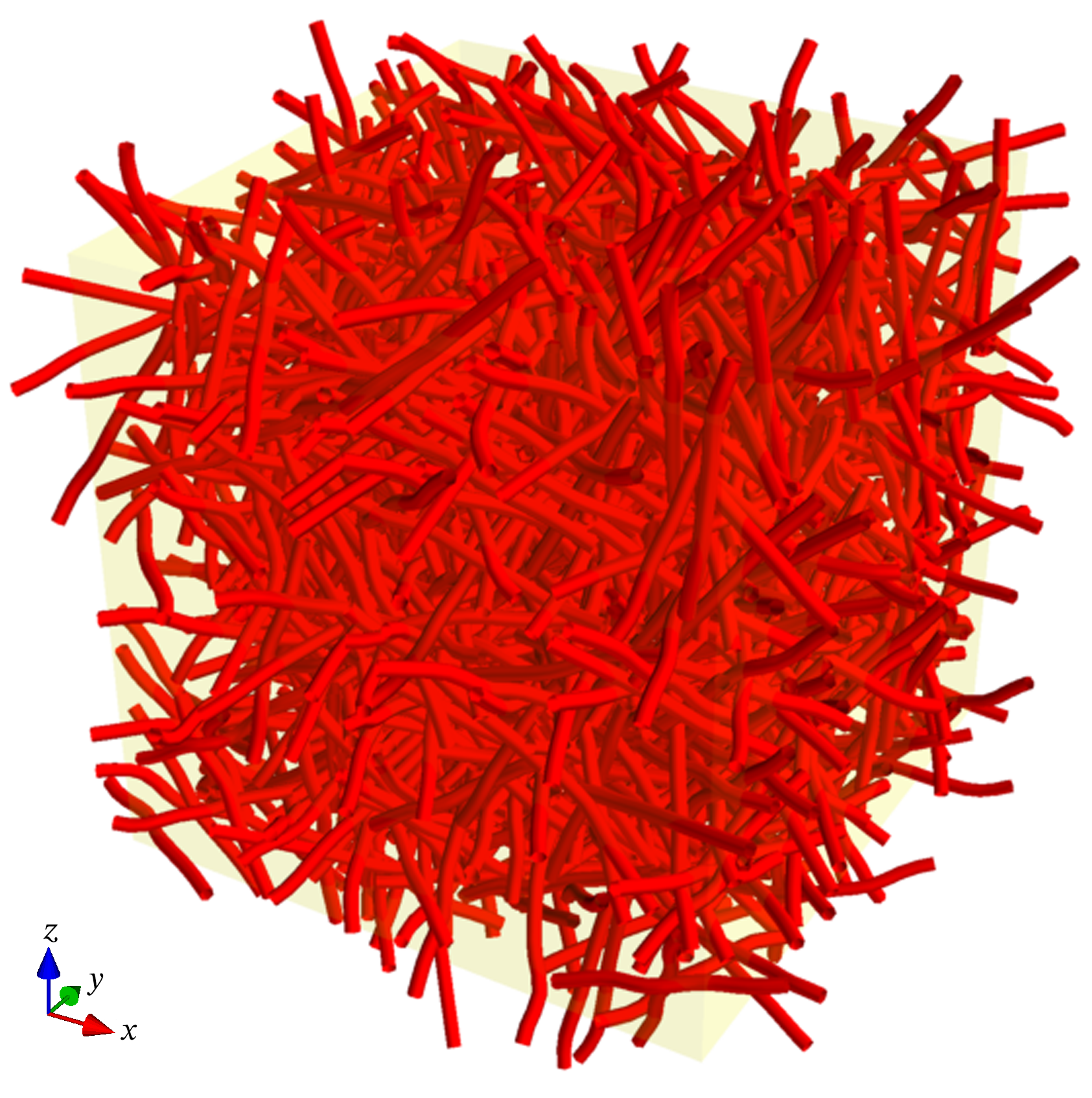}};
        \node[anchor=north west,inner sep=0] at (0,0) {(\textit{a})};
        \node at (0,-5.5) {$\;$};
    \end{tikzpicture}
    \begin{tikzpicture}
        \node[anchor=north west,inner sep=0] (image) at (0,-0.3) {\includegraphics[width=0.32\textwidth]{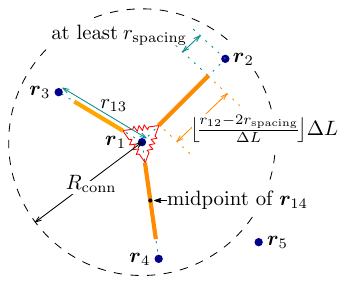}};
        \node[anchor=north west,inner sep=0] at (0,0) {(\textit{b})};
        \node at (0,-5.5) {$\;$};
    \end{tikzpicture}    
    \begin{tikzpicture}
        \node[anchor=north west,inner sep=0] (image) at (0,0) {\includegraphics[width=0.32\textwidth]{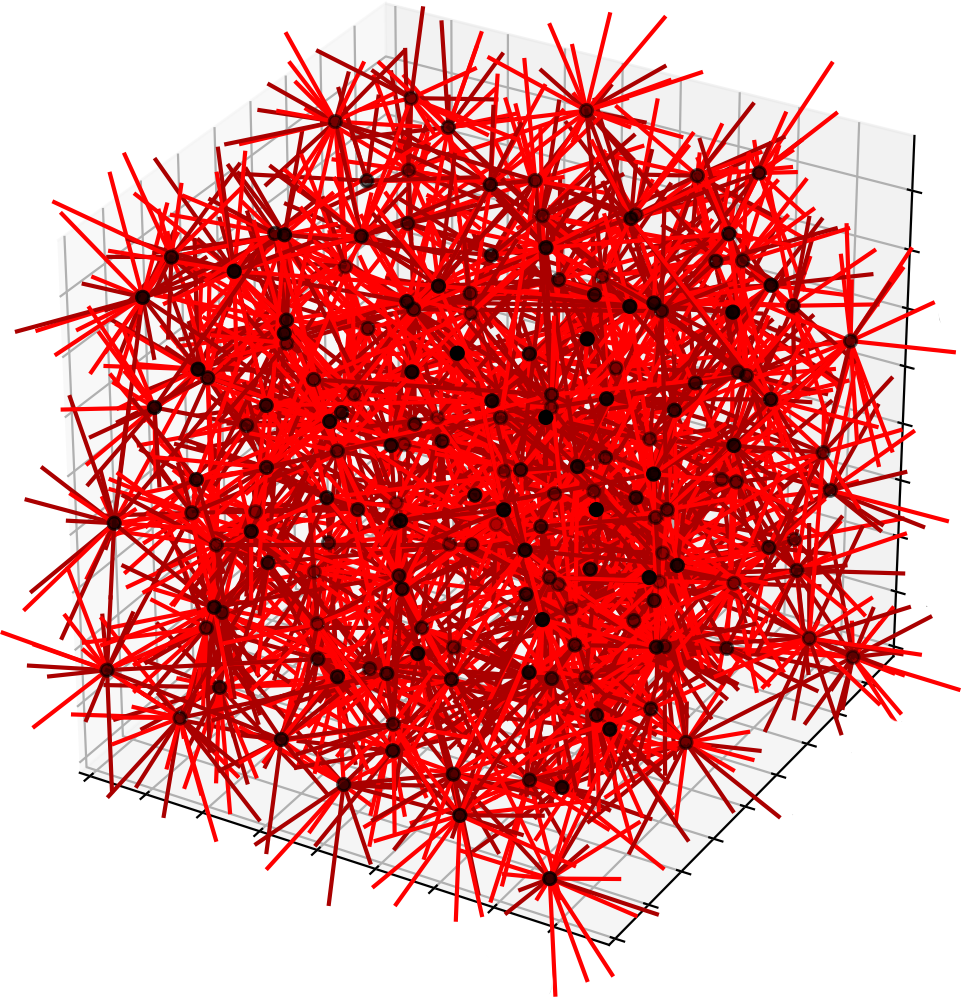}};
        \node[anchor=north west,inner sep=0] at (0,0) {(\textit{c})};
    \end{tikzpicture}        
  	\caption{Filament suspension seeding. (\textit{a}) Render of an unconnected suspension from \cref{sec:soup-seeding} in a yellow periodic box at 11.3\% volume concentration. (\textit{b}) Diagram of filaments connected by springs at the node at $\v{r}_1$, as used in \cref{sec:network-seeding}. (\textit{c}) A render of one periodic cell of the connected network suspension highlighting the placement of the connecting nodes.}
     \label{fig:seeding}
\end{figure}

Producing the filament suspension follows a common and straightforward algorithm.  We first place an undeformed swimmer (length $L=33a$) in the periodic domain.  We then randomly seed (in position and orientation) $M$ undeformed filaments of length $L=33a$.  This will inevitably lead to overlapping between filaments, which must be addressed before the simulation is run.  For low filament volume fraction, $\phi < 5\%$, where $\phi = 4 \upi a^3 L M / (3V\Delta L)$, overlapping is resolved as part of the seeding process itself, where filaments are checked for overlapping as they are seeded and the seed is rejected if overlapping occurs.  This is continued until all $M$ filaments are successfully introduced to the domain.  For high volume fractions, $\phi > 5\%$, filaments are placed at random positions and orientations in the domain, regardless of any overlap.  To remove overlapping, the filaments are allowed to evolve under very short-ranged repulsive forces with segment mobility based on Stokes drag.  The filaments are allowed to evolve until all overlapping is removed and this final configuration is used as the initial condition for our simulations.   An example of a suspension simulation initial condition for $\phi = 11.2\%$ is shown in \cref{fig:seeding}(\textit{a}).
        
Suspension simulations are conducted using two values of the nondimensional relative bending modulus, $K_B' \coloneqq K_B/K_B^\text{swim} = 10^{-3}$ and $1$, where $K_B^\text{swim}$ is the bending modulus of the swimmer.  These values correspond to relatively flexible and stiff networks, respectively. We use nondimensional parameters in our study, allowing us to transcend a link to any specific organism and environment, but we briefly present dimensional measurements from the literature for context. For \emph{C. elegans}, estimates of $K_B^\text{swim}$ range widely, from $O(10^{-16})$ to $O(10^{-13})\,\mathrm{N\,m}^2$  \citep{sznitman_material_2010,fang-yen_biomechanical_2010,backholm_viscoelastic_2013}. For mammalian sperm cells, the estimated range is at least $O(10^{-21})$ to $O(10^{-19})\,\mathrm{N\,m}^2$ \citep{rikmenspoel_tail_1965,lindemann_counterbend_2005,gadelha_mathematical_2012}. Estimates for mucin fibres are scarce in the literature, partly due to significant variation in their diameter, which ranges from 3 to 10\,nm \citep{shogren_role_1989}. Given that $K_B$ scales with the fourth power of the diameter for long, thin tubes, this results in a variation spanning four orders of magnitude. Assuming a typical biopolymer Young's modulus of $1\,\mathrm{GPa}$, the estimated $K_B$ ranges from $O(10^{-25})$ to $O(10^{-22})\,\mathrm{N\,m^2}$. 

Simulations are also performed for six different filament volume fractions between $\phi = 1.9\%$ and $18.9\%$. Videos of simulations with $\phi=7.6\%$ for both $K'_B=10^{-3}$ and $1$ are included in the supplementary material.  

\subsection{Swimming speeds in filament suspensions}
\label{sec:soups-results}

We start by recording and measuring the swimming speed of the undulating filament as it moves through the filament suspensions. For each value of $K'_B$ and $\phi$, we perform 10 independent simulations, each with different randomly generated initial filament configurations, run for 20 periods of undulation.  As the active filament swims, its stroke adds periodic contributions to the swimmer's speed. To remove these contributions, the swimming speed is defined as the period-average of the centre of mass velocity in the direction of the period-averaged swimmer orientation \citep{kamal_enhanced_2018}. 

At any given time $t$, the instantaneous velocity of the centre of mass of the swimmer is given by
    \begin{equation}
        \v{V}_{\text{CoM}} = \frac{1}{N}\sum_{n=1}^N \v{V}_n,
        \label{Vswim-def-1}
    \end{equation}
    where $\v{V}_n$ is the velocity of segment $n$. The instantaneous orientation vector of the swimmer is given by $\uv{p} = \v{p}/|\v{p}|$ where
    \begin{equation}
        \v{p} = -\frac{1}{N}\sum_{n=1}^N \uv{t}_n,
        \label{Vswim-def-2}
    \end{equation}
    and where $\uv{t}_n$ is the tangent vector of segment $n$. We can then calculate the period-averaged velocity of the swimmer, $\v{V}_\text{period}(t)$, for time $t\geq T$, as
    \begin{equation}
        \v{V}_\text{period}(t) = \frac{1}{T}\int_{t-T}^{t}\v{V}_{\text{CoM}}(t')\,\d t',
        \label{Vswim-def-3}
    \end{equation}
    while the period-averaged orientation is $\uv{P} = \v{P}/|\v{P}|$, where
    \begin{equation}
        \v{P}(t) = \frac{1}{T}\int_{t-T}^{t}\uv{p}(t')\,\d t'.
        \label{Vswim-def-4}
    \end{equation}
    The swimming speed, $V_{\text{swim}}(t)$, at any given time is then defined as
    \begin{equation}
        V_{\text{swim}}(t) = \v{V}_\text{period}(t)\cdot\uv{P}(t).
        \label{Vswim-def-5}
    \end{equation} 

\begin{figure}
        \centering
        \begin{tikzpicture}
        \node[anchor=north west,inner sep=0] (image) at (0,0) {\includegraphics[width=\textwidth]{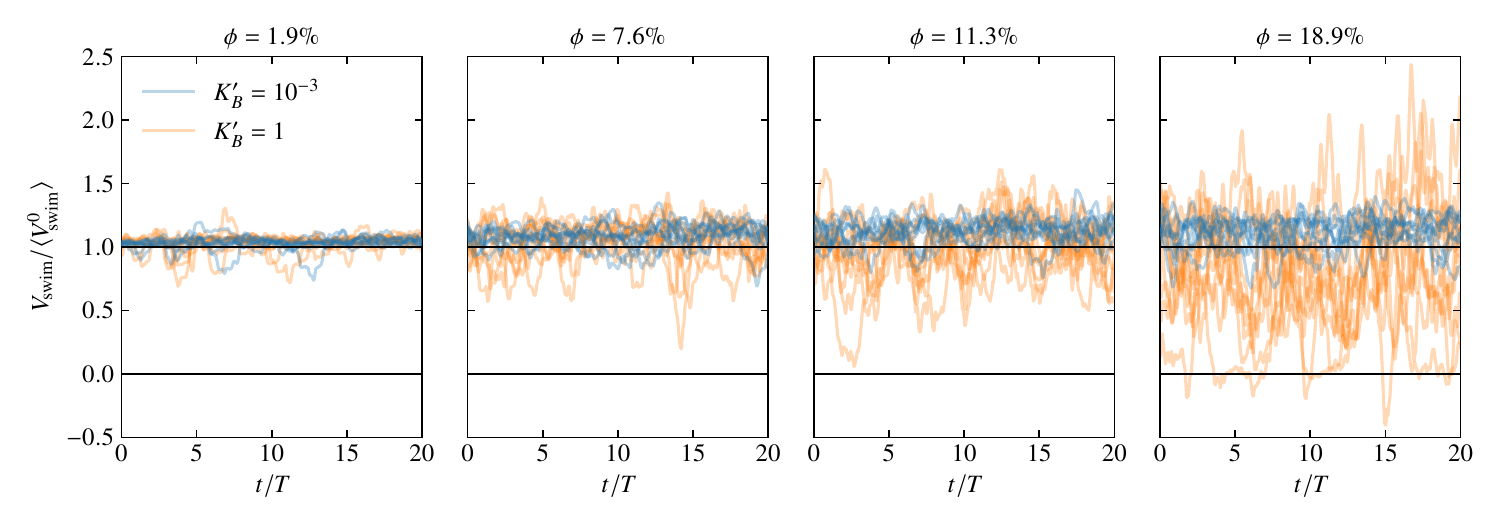}};
        \node[anchor=north west,inner sep=0] at (0,-0.2) {(\textit{a})};
        \end{tikzpicture}
        \begin{tikzpicture}
        \node[anchor=north west,inner sep=0] (image) at (0,0) {\includegraphics[width=\textwidth]{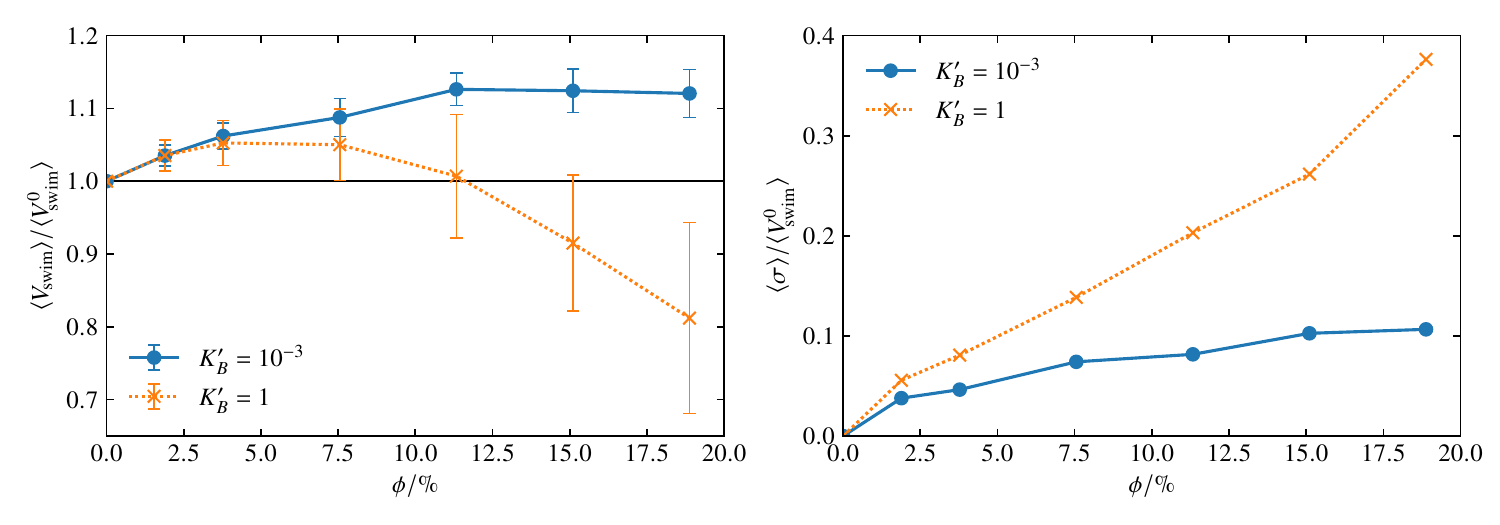}};
        \node[anchor=north west,inner sep=0] at (0,-0.2) {(\textit{b})};
        \node[anchor=north west,inner sep=0] at (8.6,-0.2) {(\textit{c})};
        \end{tikzpicture}        
        
        \caption{Swimmer speeds in the filament suspensions. (\textit{a}) Superimposed swimmer speeds over 20 periods of undulation for 10 independent simulations for each parameter pair $(\phi, K'_B)$ in a filament suspension. 
        (\textit{b}) Mean swimming speed, averaged over time and simulations, for each filament stiffness, $K'_B$, as a function of filament volume fraction, $\phi$. Error bars signify $\pm1$ standard deviation of simulations' mean over time. (\textit{c}) Mean of simulations' standard deviation, $\sigma$, over time.}
        \label{fig:swimmer-speed-over-time}
        \label{fig:mean-speed-and-std-over-phi}
        
    \end{figure}
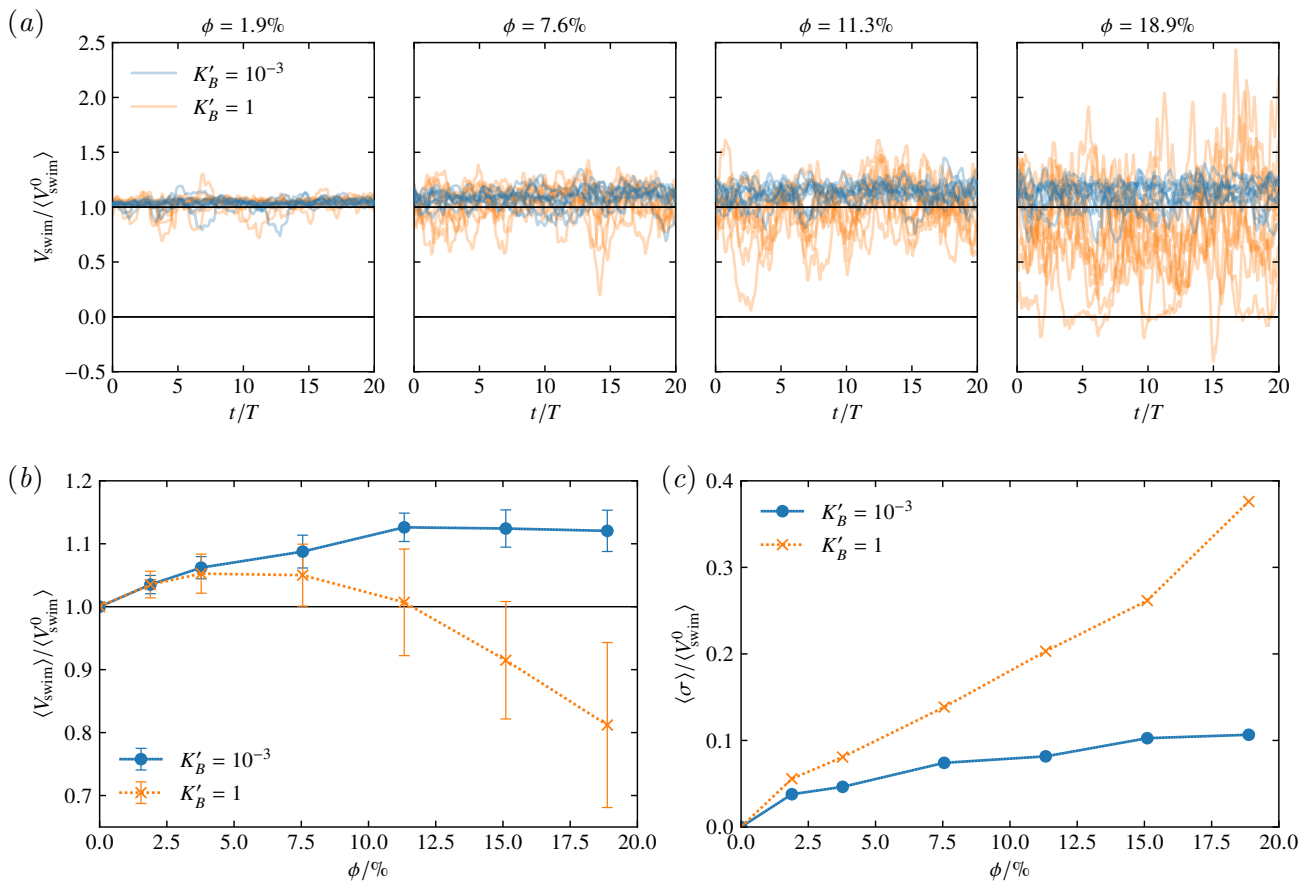

In \cref{fig:swimmer-speed-over-time}(\textit{a}), we present the normalised swimming speed $V_{\text{swim}}(t) / \langle V_{\text{swim}}^0 \rangle$, where $\langle V_{\text{swim}}^0 \rangle$ is the swimming speed in the absence of filaments, for each independent simulation, for increasing concentration and both values of the bending modulus.  Along with this, we show in panel (\textit{b}) of the same figure, $\langle V_{\text{swim}} \rangle/\langle V_{\text{swim}}^0 \rangle$, the average value of the swimming speed across both time and independent simulations.  We see that in suspensions of relatively flexible filaments ($K'_B = 10^{-3}$), the speed of the swimmer is promoted up to 15\% due to the presence of the filaments, with more concentrated suspensions yielding higher speeds.  For suspensions with stiffer filaments ($K'_B=1$), a small speed promotion of up to 5\% is observed at low concentrations, but at higher concentrations (above $\phi\sim12\%$), the swimmer is on average hindered, displaying speed decreases of up to $20\%$.
    
The speed enhancement at low concentrations is consistent with the speed increases of up to 20\% seen in simulations of two-dimensional undulatory swimmers in Stokes--Oldroyd-B viscoelastic fluids \citep{teran_viscoelastic_2010,thomases_mechanisms_2014}. 
In common with the planar undulatory motion in \citet{kamal_enhanced_2018}, the average swimmer slowdown for rigid, concentrated networks is not due to a uniform speed decrease across all independent runs; instead, it is due to large variations in the speed, including periods of time where the swimmer is trapped within the network in some simulations. 
But in contrast, our speed enhancement is significantly less than in simulations of interrupted planar swimming, where fixed \citep{majmudar_experiments_2012} or tethered \citep{kamal_enhanced_2018} discrete obstacles lead to three- to tenfold speed increases. We also see less enhancement than in helical swimming with fixed obstacles (at 40\%: \citealp{leshansky_enhanced_2009}), where swimming speed can increase monotonically with concentration \citep{klingner_helical_2020}. We discuss mechanisms for these distinctions in \cref{sec:discussion}.
    
The fluctuations in the swimmer speed seen in \cref{fig:swimmer-speed-over-time}(\textit{a}) increase with concentration, but this effect is much larger for the stiffer networks. At the highest concentration, the swimmer is sometimes sped up by a factor of 2, but also sometimes forced \emph{backwards} in a number of simulations.  This is quantified by the standard deviation in panel (\textit{c}) of the same figure, where the mean standard deviation, $\langle\sigma\rangle$, increases almost linearly with concentration and reaches $0.4$ of the unhindered swimmer speed in the most concentrated suspensions of stiff filaments.

\section{Spring-connected filament networks}
\label{sec:networks}

\subsection{Filament environment}
\label{sec:network-seeding}
\begin{figure}
    \centering
    \begin{tikzpicture}
        \node[anchor=north west,inner sep=0] (image) at (0,0) {\includegraphics[width=\textwidth]{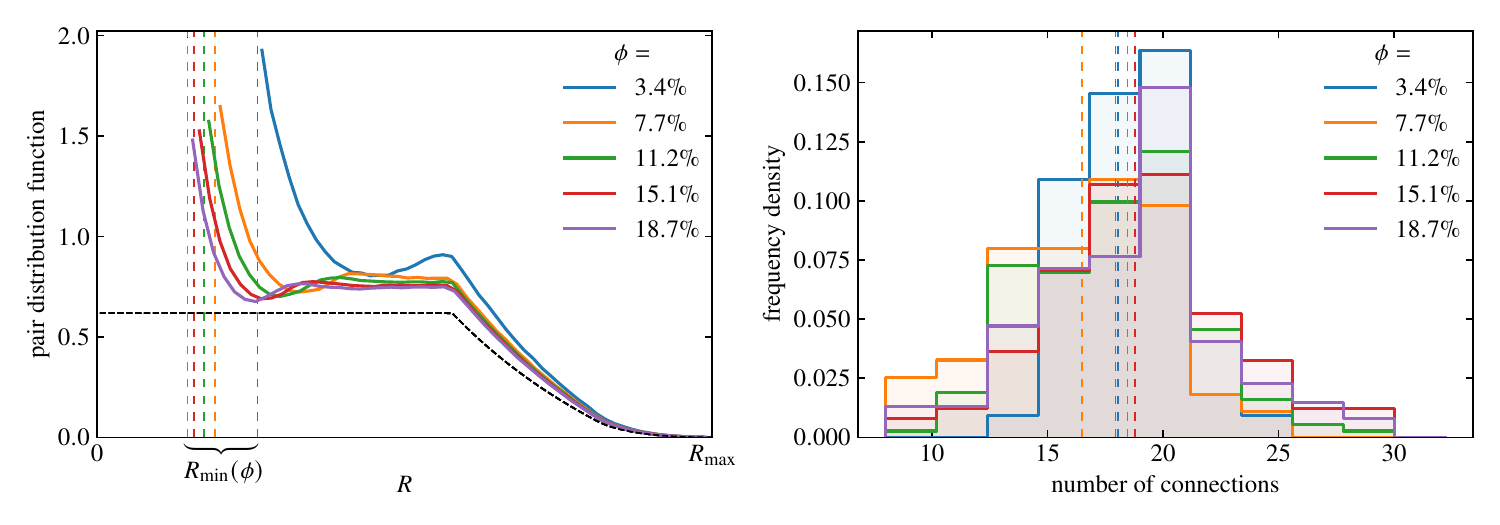}};
        \node[anchor=north west,inner sep=0] at (0,-0.2) {(\textit{a})};
        \node[anchor=north west,inner sep=0] at (8.6,-0.2) {(\textit{b})};
    \end{tikzpicture}   
    \caption{Spring network seeding statistics. (\textit{a}) Pair distribution function of node spacing for each of our concentrations, averaged over 1000 trials. The horizontal black dotted line corresponds to the pair distribution function of a uniform distribution in a periodic domain \citep{deserno_how_2004}.
    (\textit{b}) Histogram of the number of connections at each node for each of our concentrations, averaged over 100 trials. The mean in each case, represented by the vertical dotted line, is between 16 and 19.}
    \label{fig:g-of-r-network}
    \label{fig:distribution-of-connections-network}
\end{figure}

\begin{table}
  \begin{center}
    \begin{tabular}{llccccc}
    \toprule
         Filament volume concentration & $\phi$ & 3.4\% & 7.7\% & 11.2\% & 15.1\% & 18.7\%  \\
    \midrule
         Number of nodes seeded & $N_\text{nodes}$ & 50 & 125 & 170 & 225 & 280 \\
         Minimum node separation & $R_\text{min} = 2.2LN^{-1/3}_\text{nodes}$ & $0.60L$ & $0.44L$ & $0.40L$ & $0.36L$ & $0.34L$\\
         Connect nodes if closer than & $R_\text{max}$ & $1.2L$ & $1.2L$ & $1.2L$ & $1.2L$ & $1.2L$\\
         Connect with probability & $P_\text{conn}$ & 1 & 0.35 & 0.28 & 0.22 & 0.17\\
    \bottomrule
    \end{tabular}
    \caption{Spring network seeding parameters for each concentration $\phi$, used in the algorithm in \cref{sec:network-seeding}.}
    \label{tab:network-statistics}
    \end{center}
\end{table}

A number of methods for generating linked filament networks exist in the literature. One approach, seen in two-dimensional \citep{head_deformation_2003, head_distinct_2003, didonna_filamin_2006, heussinger_stiff_2006} and three-dimensional \citep{astrom_strain_2008} models of actin networks, is to seed straight filaments isotropically and add crosslinks where filaments intersect, potentially removing loose ends \citep{wilhelm_elasticity_2003, huisman_three-dimensional_2007}. Another option is to seed straight filaments, but let them grow into other shapes according to an algorithm, before forming crosslinks \citep{buxton_actin_2009}. 

Here, we use a probability-based algorithm to construct a phenomenological model of a generic network, based on \citet{buxton_bending_2007}. We place the swimmer, of length $L$, in the empty periodic domain and seed $N_{\text{nodes}}$ nodes randomly inside the domain at a distance at least $R_\text{min} = 2.2LN_{\text{nodes}}^{-1/3}$ from each other. The exponent of $-1/3$ corresponds to hard sphere packing inside a finite domain.  For each pair of nodes, $i$ and $j$, with positions $\v{r}_i$ and $\v{r}_j$, respectively, we place a filament connecting these nodes with a probability $P_{\text{conn}}$ if the distance between nodes is $|\v{r}_i - \v{r}_j| = |\v{r}_{ij}| < R_{\text{max}} = 1.2L$.  The filament is placed along the vector $\v{r}_{ij}$, with its centre at the midpoint of $\v{r}_{ij}$ and formed of $N=\lfloor (r_{ij}-2r_{\text{spacing}})/\Delta L \rfloor$ segments, where $r_{\text{spacing}} = 2a$ is taken to ensure filaments do not overlap at the nodes.  Linear spring forces $-k_s(x-\ell)\uv{x}$, where $\v{x}=x\uv{x}$ is the end-to-end vector and $\ell$ is the natural spring length, are applied between all pairs of filament end segments that meet at a node. In doing this, we generate a network spanning the periodic domain with no external force holding it in place; in practice, the drag on the component filaments limits the network from moving as a rigid body. The seeding algorithm is summarised in \cref{fig:seeding}(\textit{b}).
       
To ensure that filaments do not overlap, this seeding procedure is carried out in a separate simulation, using friction-based mobility and an active collision barrier, until an equilibrium has been found. We once again define the volume concentration of the network by treating each segment as a sphere with its hydrodynamic radius $a$. That is,  for $M$ filaments of length $L_i$ in a periodic domain with volume $V$, we have $\phi = 4 \upi a^3 \sum_{i=1}^M{L_i} / (3V\Delta L)$. Under this definition, the concentration scales as the square of the number of nodes, $\phi \sim N_{\text{nodes}}^2$. Over a large number of independent random initial seeds, the concentration is normally distributed with a standard deviation that grows linearly with $N_{\text{nodes}}$. 

Simulations are performed for different concentrations of filaments between $\phi = 3.4\%$ and $18.7\%$. We choose $N_\text{nodes}$ and $P_\text{conn}$ to produce networks of a given concentration with a common mean filament length of approximately $0.75L$, and common mean number of connections at each node (to within a reasonable tolerance). Our parameter choices are summarised in \cref{tab:network-statistics}.
    
The pair distribution function of node separations under these configurations is displayed in \cref{fig:g-of-r-network}(\textit{a}). We see a bump near $R_\text{min}$ corresponding to close packing, but otherwise the PDF follows the shape for a uniform distribution, slightly enhanced. A render of one such network at 11.3\% volume concentration is shown in \cref{fig:seeding}(\textit{c}), with the nodes represented by black circles.
    
Simulations are performed for two different values of the network filament bending modulus, $K'_B = 10^{-3}$ and $1$ (as for the filament suspensions), and for three values of the spring modulus, $k'_s = 0,0.36,3.6$, where $k'_s \coloneqq k_sL^3/K_B^\text{swim}$. The swimmer bending and twist modulus remains constant, and the twist modulus, $K_T$, for every filament, is set equal to its bending modulus.
    
Simulations of flagellar swimming through a viscoelastic network formed of virtually cross-linked nodes by \citet{wrobel_enhanced_2016} show that the connectivity number matters: higher-connectivity networks are stiffer. Here, we keep this number constant so that we can explore the effect of the other parameters in the system: the network filament bending modulus, the filament concentration, and the spring modulus of the networks. The distribution of connection numbers in each case is shown in \cref{fig:distribution-of-connections-network}(\textit{b}), where the mean connectivity number in each of our simulations is between 16 and 19. For comparison, cubic lattices with the diagonals excluded and included have connectivity numbers of 6 and 26 respectively, and were used in \citet{wrobel_enhanced_2016}. \Citet{buxton_bending_2007} use connectivity numbers between 5 and 15.

\subsection{Swimming speeds in filament networks}
\label{sec:networks-results}

We now present results from simulations examining swimmer motion in the connected filament networks.  We again explore the effects of filament bending stiffness and volume fraction, but now also include the spring constant of the connections. For each triplet of parameters $(K'_B,\phi,k'_s)$, five simulations are performed, each with a different initial random configuration, and each running for 20 undulation periods.
    
\begin{figure}
    \centering
    \includegraphics[width=\textwidth]{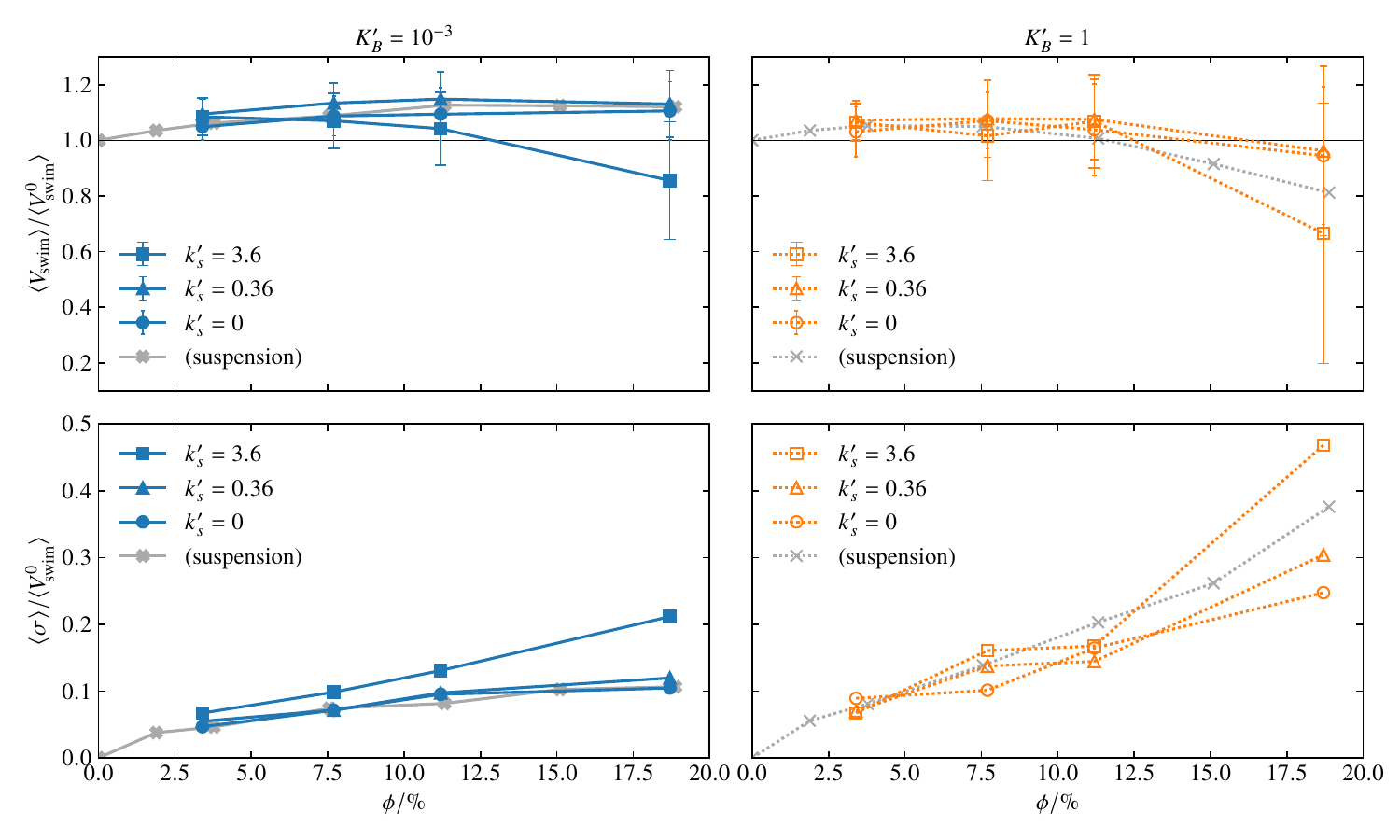}
    \caption{Swimmer speeds in the filament networks. Mean normalised swimmer speed and standard deviation over 20 periods of undulation, as a function of the volume fraction of network filaments, for both values of the network filament bending modulus, and for all three values of the network spring strength $k'_s$. Results for each parameter set are averaged from 5 independent simulations, and error bars signify $\pm1$ standard deviation over the 5 simulations. For comparison, the suspension measurements from \cref{fig:mean-speed-and-std-over-phi} are included in pale grey.}
    \label{fig:mean-speed-and-std-over-phi-network}
\end{figure}    

The mean speed and standard deviation over the simulation run, $\langle V_\text{swim} \rangle$ and $\langle \sigma \rangle$, are presented in \cref{fig:mean-speed-and-std-over-phi-network}.  
We perform simulations for three different values of the spring constant.  In the case where $k'_s=0$, the simulations differ from the filament suspensions in \cref{sec:soups} only in the seeding algorithm. For comparison, $\langle V_\text{swim} \rangle$ and $\langle \sigma \rangle$ from the suspension simulations are reproduced in this figure.  We see that for $k'_s=0$ and $0.36$, the trend in the data follows that found for filament suspension simulations: swimmer speed is promoted with concentration in flexible networks ($K'_B=10^{-3}$) up to $\sim15\%$, but rigid networks ($K'_B=1$) produce a small speed promotion followed by a hindrance as concentration increases. 

The more striking differences with the suspensions simulations are observed when $k'_s=3.6$, where the filament tethers are stiffer than the swimmer.  Here, the swimmer speed in the flexible network decreases monotonically with concentration, down to an approximate reduction of $12\%$. The swimmer is hindered even further in the stiff filament network, with speeds decreasing by approximately $35\%$ at the highest concentrations.  Interestingly, at intermediate concentrations, there is little difference with the weaker-spring scenarios. The effects of the stronger spring constant also yield considerably higher fluctuations in swimming speed with a standard deviation of up to $0.2\langle V_{\text{swim}}^0 \rangle$, although notably not as high as in the unconnected suspensions from \cref{sec:soups} where the standard deviation can be as high as $0.4 \langle V_{\text{swim}}^0 \rangle$.

\section{Investigation into mechanisms for motility changes}

In the previous sections, we observed that the swimming speed is mildly promoted by the presence of filaments at low concentrations: the mere existence of the filaments (either flexible or stiff) in suspension produces this effect.  The addition of weak connections between the filaments can provide an additional small speed increase across all concentrations and all rigidities.  However, for cases with stiffer filaments or stiffer connecting springs, high concentrations of filaments tend to slow the swimmer down.  

In this section, we aim to explore the mechanisms at play by which swimmer speed is promoted or hindered.  Specifically we seek to gain insight into these trends by examining the forces acting on the swimmer and changes in swimming gait due to the presence of the filaments. The repulsive barrier forces act on the swimmer as it passes through the suspension or network of filaments. These forces act to oppose contact with suspended filaments and can therefore act on the swimmer in all directions.  We can consider these forces to act in two main ways: (i) by considering the total force acting on the swimmer, and (ii) by considering how the shape of the swimmer changes as a result of unequal forces along the swimmer body.

\subsection{The role of steric forces directly}
The barrier force can result in a nontrival net force on the swimmer that can push or pull it through the environment. We compute the mean period-averaged barrier force in the swimming direction, $\langle F_\text{swim} \rangle$.  This can be measured analogously to $\langle V_{\text{swim}}\rangle$ in \cref{Vswim-def-1,Vswim-def-2,Vswim-def-3,Vswim-def-4,Vswim-def-5}, namely
    \begin{equation}
        \v{F}_\text{total} = \sum_{n=1}^N \v{F}_n^B, \quad \v{F}_\text{period}(t) = \frac{1}{T}\int_{t-T}^t\v{F}_\text{total}(t')\,\d t', \quad F_\text{swim}(t) = \v{F}_\text{period}(t)\cdot\uv{P}(t),
    \end{equation}
noting that since the constraint forces on the swimmer sum to zero, the total barrier force is equal to the total hydrodynamic force, $\sum_n \v{F}_n^B = \sum_n \v{F}_n^H$.

\begin{figure}
    \centering
    \begin{tikzpicture}
        \node[anchor=north west,inner sep=0] (image) at (0,0) {\includegraphics[width=.85\textwidth]{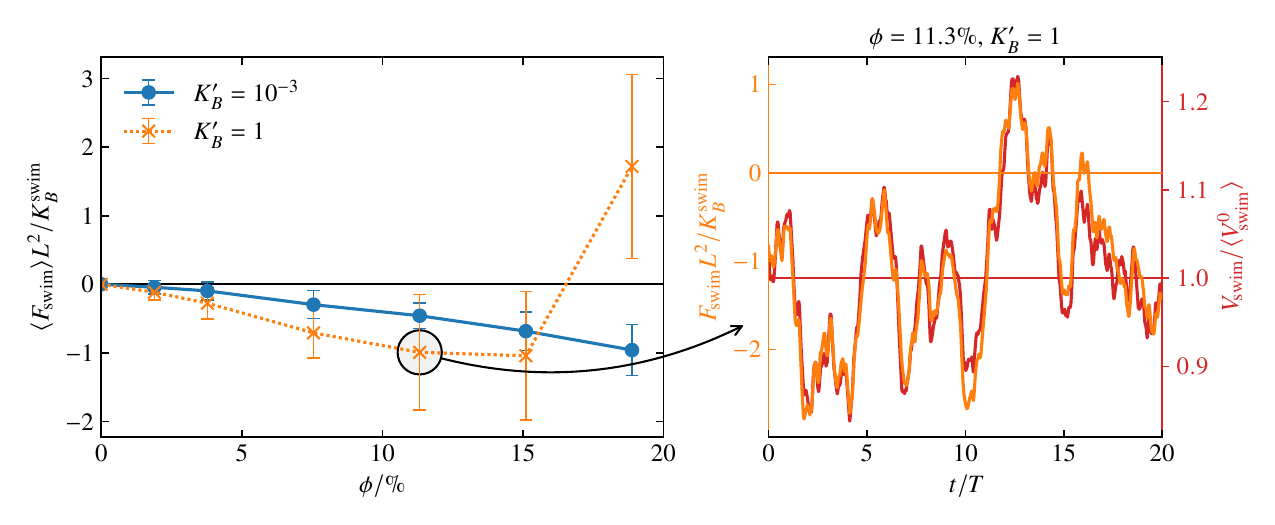}};
        \node[anchor=north west,inner sep=0] at (0,-0.4) {(\textit{a})};
        \node[anchor=north west,inner sep=0] at (8.0,-0.4) {(\textit{b})};
    \end{tikzpicture}       
    \caption{
    Force from the suspensions. While forces correlate with speed fluctuations, on average they push backwards (with the outlying high-$\phi$, high-$K'_B$ result indicative of a different regime we will also see in the spring networks).
    (\textit{a}) Mean period-averaged force on the swimmer as a function of $\phi$, nondimensionalised on the swimmer length and bending modulus. Error bars signify $\pm1$ standard deviation over the 10 simulations.
    (\textit{b}) Period-averaged force on the swimmer over time for the circled data point on the left, overlaid with the swimming velocity. Fluctuations in the force are seen to correlate with speed, but consider the region between the two horizontal bars: the swimmer speed often remains above the swimming speed in the absence of filaments (red line), despite the forces being negative, i.e.\ pushing backwards (orange line).}
    \label{fig:forces-cause-fluctuations}
    \label{fig:forces-push-back}        
\end{figure}
    
The total force, $\langle F_\text{swim} \rangle$, in the suspensions is presented in \cref{fig:forces-push-back}(\textit{a}). For all but the most concentrated case with the stiff filaments, the steric forces act against the direction of swimming. This is somewhat surprising as everywhere where this force is opposite the swimming direction, the velocity plot in \cref{fig:mean-speed-and-std-over-phi} showed us that the swimmer is travelling \emph{faster} than it does in the absence of filaments. The high-concentration, high-stiffness case appears an outlier, being pushed on average forwards, indicative of a different regime which will appear again in the spring networks. 
    
Although the direction of the mean force appears at odds with the increase in swimming speed, an examination of the time-series before averaging shows a direct correlation between force and speed \emph{fluctuations}, as shown in \cref{fig:forces-cause-fluctuations}(\textit{b}).  But, still, the region between the horizontal lines in this panel represents times when the forces oppose swimming, and yet the swimmer is still travelling faster than it would in the absence of filaments, suggestive of a uniform promotion in speed across the simulation.
    
\begin{figure}
    \centering
    \includegraphics[width=\textwidth]{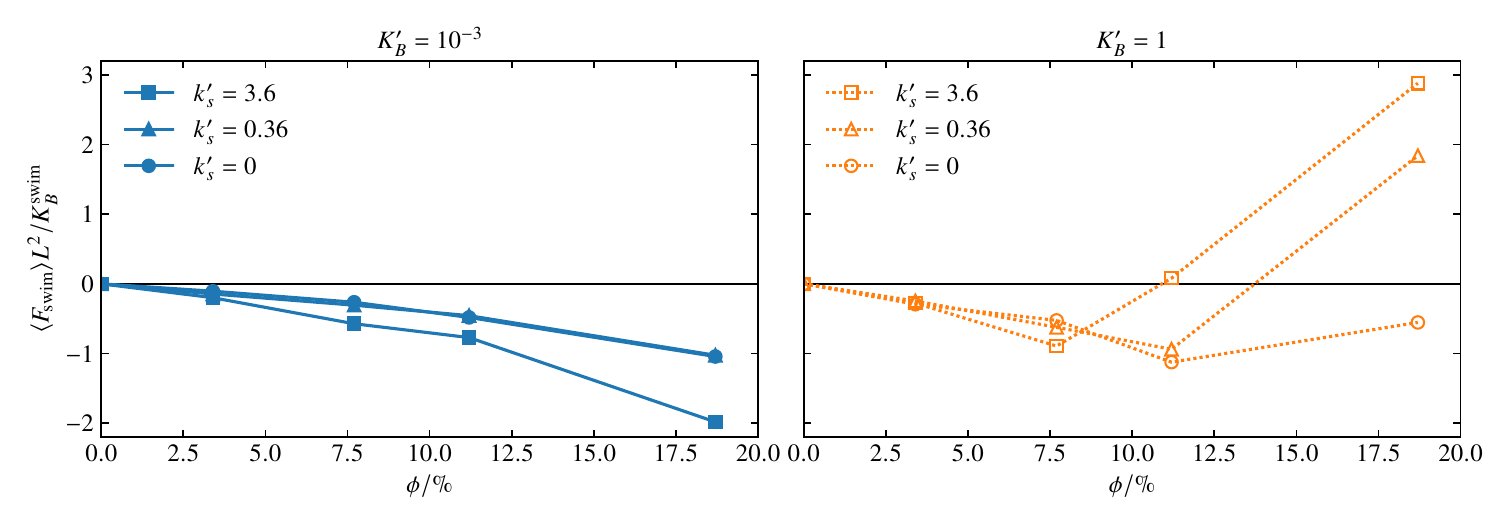}
    \caption{Forces from the spring networks: Mean force on the swimmer in the swimming direction. Plotted as a function of the volume concentration of network filaments, for both values of the network filament bending modulus and all three spring constants. Results for each parameter set are averaged from 5 independent simulations. This is analogous to \cref{fig:forces-push-back} in the filament suspensions.}
    \label{fig:forces-push-back-network}
\end{figure}
    
The filament networks follow similar trends but paint a more nuanced picture, with the total forces presented in \cref{fig:forces-push-back-network}.  As in the suspensions, the steric forces from interactions with \emph{flexible} network filaments push against the direction of swimming, with larger concentrations leading to stronger opposing forces. Once again, this happens despite the swimmer moving faster in these networks than they would at 0\% concentration (\cref{fig:mean-speed-and-std-over-phi-network}), although this isn't universal: at the highest spring constant the filament speed does drop below its filament-free value. Interestingly, increasing the spring constant from $k'_s=0$ to $0.36$ makes little difference to the force yet mildly promotes swimming speed.  At low filament concentrations, the swimmer behaves similarly in stiff and flexible filament networks, however, we see that above a critical concentration the forces from stiff networks will push in the direction of swimming yet the swimmer slows down. 
The force is much weaker for the $(k'_s=0, K'_B=1)$ case than in the matching suspension case, which we attribute to the differences in filament seeding. Increasing the spring constant recaptures and enhances the behaviour seen in the suspension.      

These results suggest that there are two force regimes for the swimmer: one where the swimmer is gently resisted, almost linearly with concentration; and one where the swimmer is pushed forwards by the network. What is surprising is that these two force regimes do not clearly translate to observable swimming speed regimes. In particular, the barrier forces cannot be responsible for the speed increase seen in suspensions at lower-to-mid concentrations. We conclude that something else is acting to mask the total applied force.

\subsection{The role of the swimming gait}

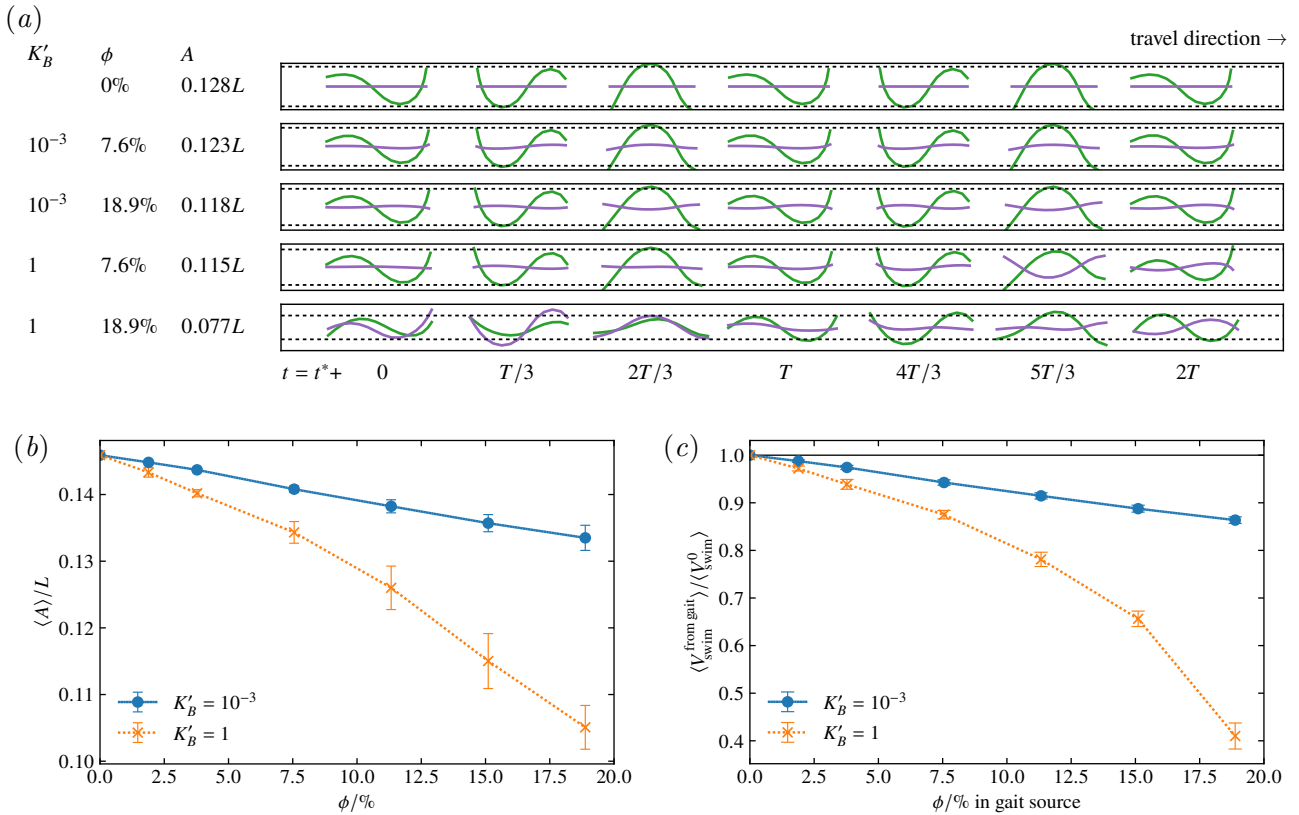
\begin{figure}
    \centering
    \begin{tikzpicture}
        \node[anchor=north west,inner sep=0] (image) at (0,0) {\includegraphics[width=\textwidth]{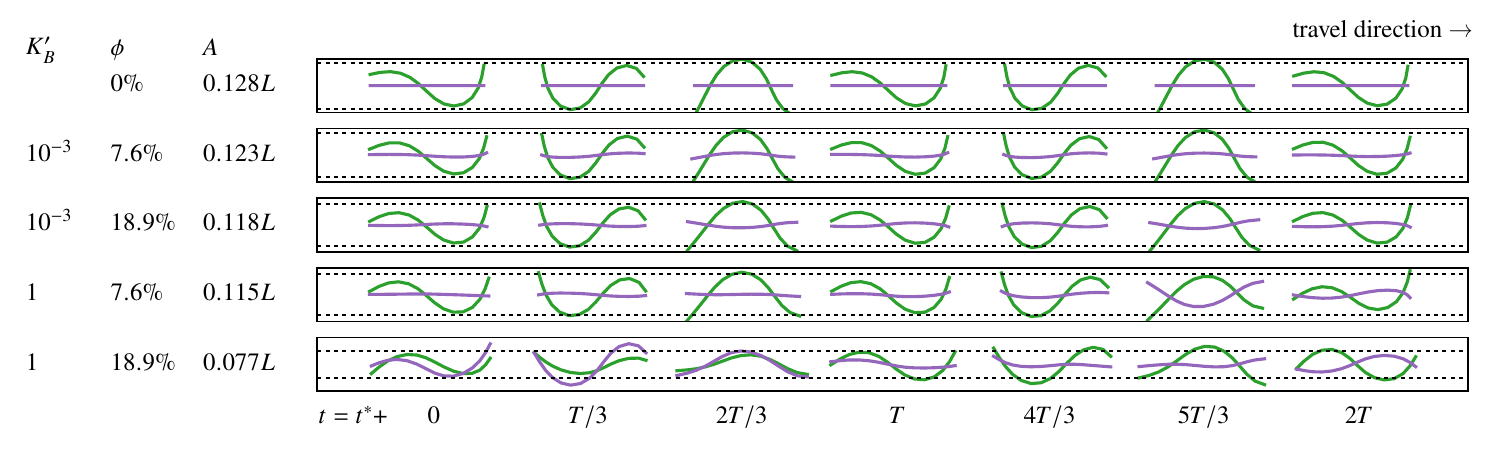}};
        \node[anchor=north west,inner sep=0] at (0,0.1) {(\textit{a})};
    \end{tikzpicture}  
    
    \vspace{2mm}
    \begin{tikzpicture}
        \node[anchor=north west,inner sep=0] (image) at (0,0) {\includegraphics[width=0.49\textwidth]{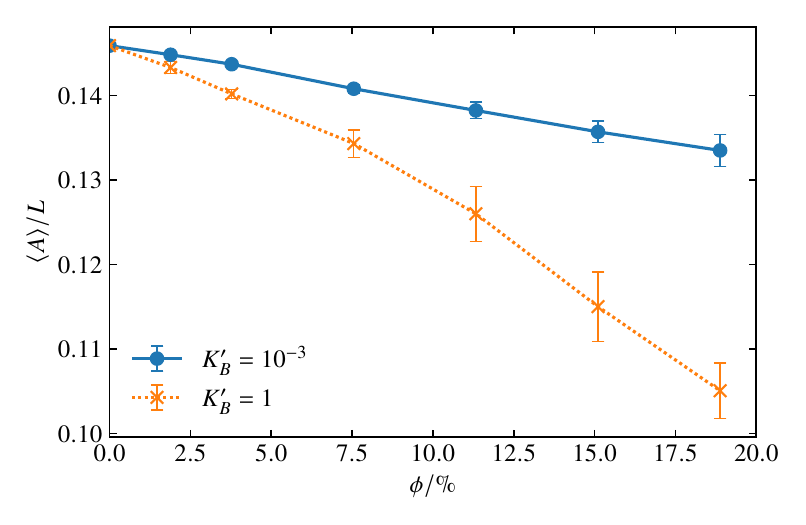}};
        \node[anchor=north west,inner sep=0] at (0,-0.2) {(\textit{b})};
    \end{tikzpicture}     
    %
    \begin{tikzpicture}
        \node[anchor=north west,inner sep=0] (image) at (0,0) {\includegraphics[width=0.49\textwidth]{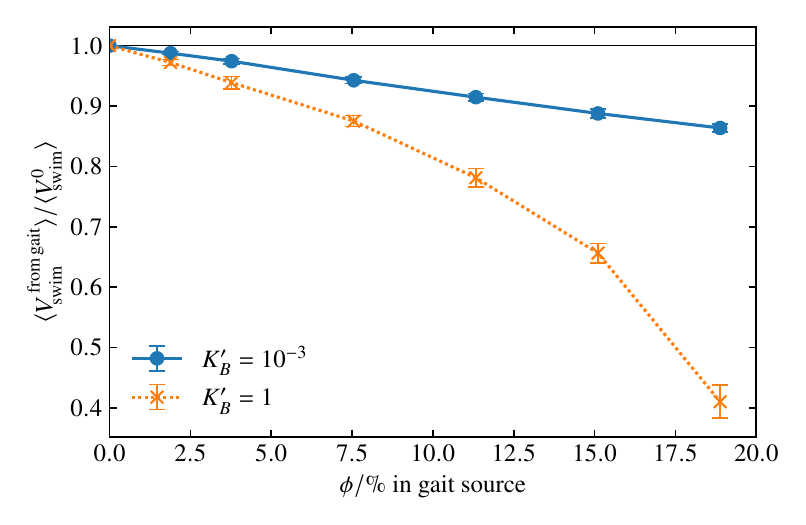}};
        \node[anchor=north west,inner sep=0] at (0,-0.2) {(\textit{c})};
    \end{tikzpicture}         
    \caption{
    Swimming gait amplitudes and effectiveness in suspensions.
    Higher concentrations lead to smaller-amplitude swimming gaits and slower swimming speeds.
    (\textit{a}) Snapshots of sample swimming gaits, moving from left to right, in the beating plane (green, `side on') and in the out-of-beating plane (purple, `top down'). Snapshots are from a suspension simulation and take place over two periods of undulation after a given time $t^*$. The mean amplitude in the swimming direction--beating plane, $A$, is marked with dotted black lines. As panel (\textit{b}) summarises, the amplitude decreases as concentration increases, which coincides with deflection in the out-of-beating plane. This is emphasised when the suspended filaments are more rigid. Videos of the $\phi=7.6\%$ cases, with the surrounding filaments visible, are included in the supplementary material.
    (\textit{b}) Mean amplitude in the beating plane, $\langle A \rangle$, for increasing concentration of suspension filaments, and for suspension filaments with bending modulus $K'_B=10^{-3}$ and $1$. The amplitude decreases with $\phi$.
    (\textit{c}) Mean speed achieved with the extracted swimmer gait when placed in a 0\% concentration fluid, $\langle V_\text{swim}^\text{from\,gait}\rangle$, subject to no external forces, following the procedure in \cref{sec:rbm}. This correlates well with the amplitude graph to the left, and shows that the gait enforced by the filament suspension is less effective at swimming than the unrestricted gait in an empty fluid. 
    In both (\textit{b}) and (\textit{c}), error bars signify $\pm1$ standard deviation over the 10 simulations.}
    \label{fig:gaits}
    \label{fig:amplitude-over-phi}
    \label{fig:gait-sucks}
\end{figure}

Along with modifying the swimmer's rigid body motion, steric forces can also act to change the swimming gait. Swimmer undulations produce collisions with the surrounding filaments, inducing lateral forces which can reduce the amplitude of the undulation and alter the beat plane, making the gait less effective.  Sample gaits from the suspension simulations, in both the beating and out-of-beating plane, are illustrated at increasing concentration in \cref{fig:gaits}(\textit{a}). These stills show that in flexible suspensions, increased concentration leads to slightly reduced beating-plane amplitudes and minor out-of-plane movements. But in rigid suspensions, especially at the highest concentration (bottom row), sample gaits deviate significantly from beating behaviour in a filament-free environment, with out-of-plane amplitude sometimes surpassing in-plane amplitude.

\begin{figure}
    \centering
    \begin{tikzpicture}
        \node[anchor=north west,inner sep=0] (image) at (0,0) {\includegraphics[width=\textwidth]{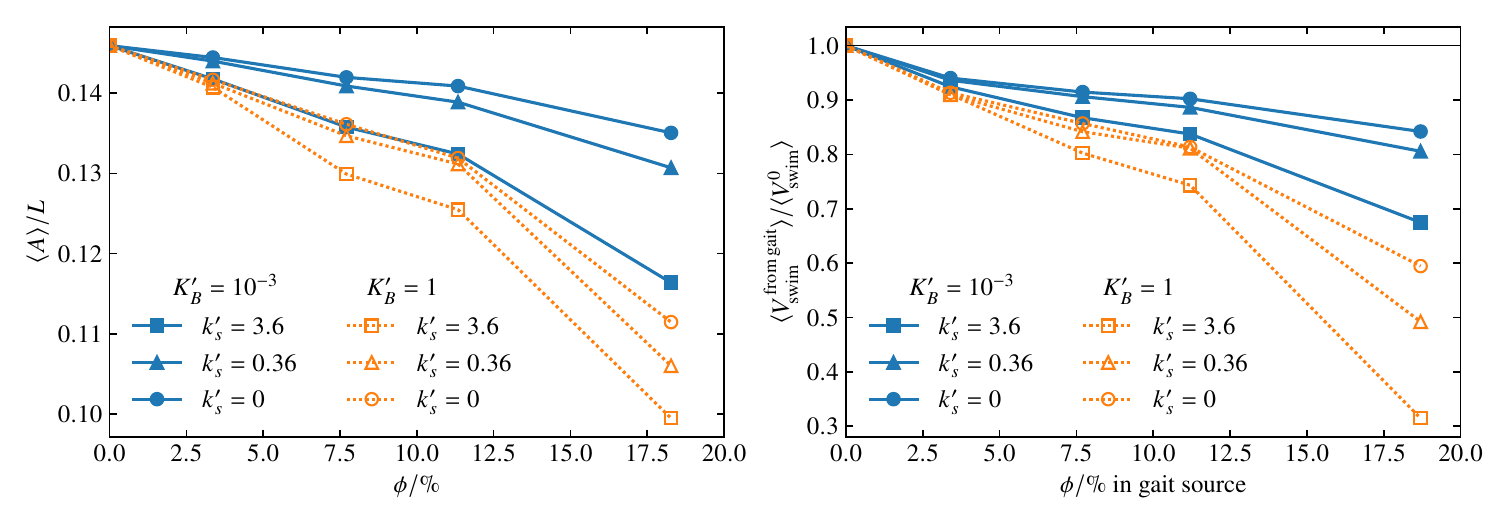}};
        \node[anchor=north west,inner sep=0] at (0,-0.2) {(\textit{a})};
        \node[anchor=north west,inner sep=0] at (8.65,-0.2) {(\textit{b})};
    \end{tikzpicture}
    \caption{Swimming gait amplitudes and effectiveness in networks. (\textit{a}) Mean amplitude in the beating plane for increasing concentration of network filaments, and for network filaments with bending modulus $K'_B=10^{-3}$ and $1$ connected with springs with three different spring constants, $k'_s$. 
    (\textit{b})
    Mean speed achieved with the extracted swimmer gait from the network simulations, following the procedure in \cref{sec:rbm}. This is the network analogue of \cref{fig:gait-sucks}. Combined with \cref{fig:forces-push-back-network}, this demonstrates that collectively, the gait and total barrier force, apart from in the most concentrated case with the stiffest filaments and spring constants, act to slow the swimmer down.}
    \label{fig:amplitude-over-phi-network}
    \label{fig:gait-sucks-network}
\end{figure} 

We can quantify these observations by measuring the mean amplitude in the beating plane, presented in \cref{fig:amplitude-over-phi}(\textit{b}) for suspensions and in \cref{fig:amplitude-over-phi-network}(\textit{a}) for our networks. We find that increased filament stiffness, $K'_B$, and for the networks, increased spring constants, $k'_s$, lead to smaller amplitudes at higher concentrations.  
        
We can assess further the efficacy of these reduced-amplitude gaits by examining how well they propel the swimming in the absence of filaments.  To do this, we remove the rigid body motion from the swimmer in the filament environment and recompute it in a viscous fluid absent of any filaments using the remaining undulations.  We describe this computation in detail in \cref{sec:rbm}.  The resulting speed when the imposed-gait swimmer swims is shown in \cref{fig:gait-sucks}(\textit{c}) and \cref{fig:gait-sucks-network}(\textit{b}). Correlating strongly with the stroke amplitude in both the suspensions and the networks, we see that the imposed gait is less effective at swimming, and increasingly so for the higher $K'_B$ and (in the network) higher $k'_s$.  Thus, any gait changes result in reduced swimming speeds rather than the increases that we observe in the full simulations.

\subsection{The role of hydrodynamics}

So far we have seen that steric interactions, at mid-to-low concentrations, hinder swimming speed directly through rigid body forces. Additionally, when we remove all suspended filaments and impose the swimming gait seen in the full simulations, the swimmer again moves with a reduced speed compared to its natural gait in an empty fluid.   

To understand the speed increases noted in \cref{fig:mean-speed-and-std-over-phi,fig:mean-speed-and-std-over-phi}, having eliminated steric forces and swimming gait, we turn our attention to hydrodynamic interactions.  To investigate this, we run our suspension simulations again, including all filaments but now without hydrodynamic interactions between the suspended bodies.  We accomplish this by solving the fluid mobility problem in \cref{eq:mobility-problem} using a resistive force theory (RFT) \citep{lauga_hydrodynamics_2009, johnson_flagellar_1979} with coefficients tuned to recover the correct filament dynamics as in \citet{schoeller_flagellar_2018}.  The details are provided in \cref{sec:rft}. 

\begin{figure}
    \centering
    \begin{tikzpicture}
        \node[anchor=north west,inner sep=0] (image) at (0,0) {\includegraphics[width=\textwidth]{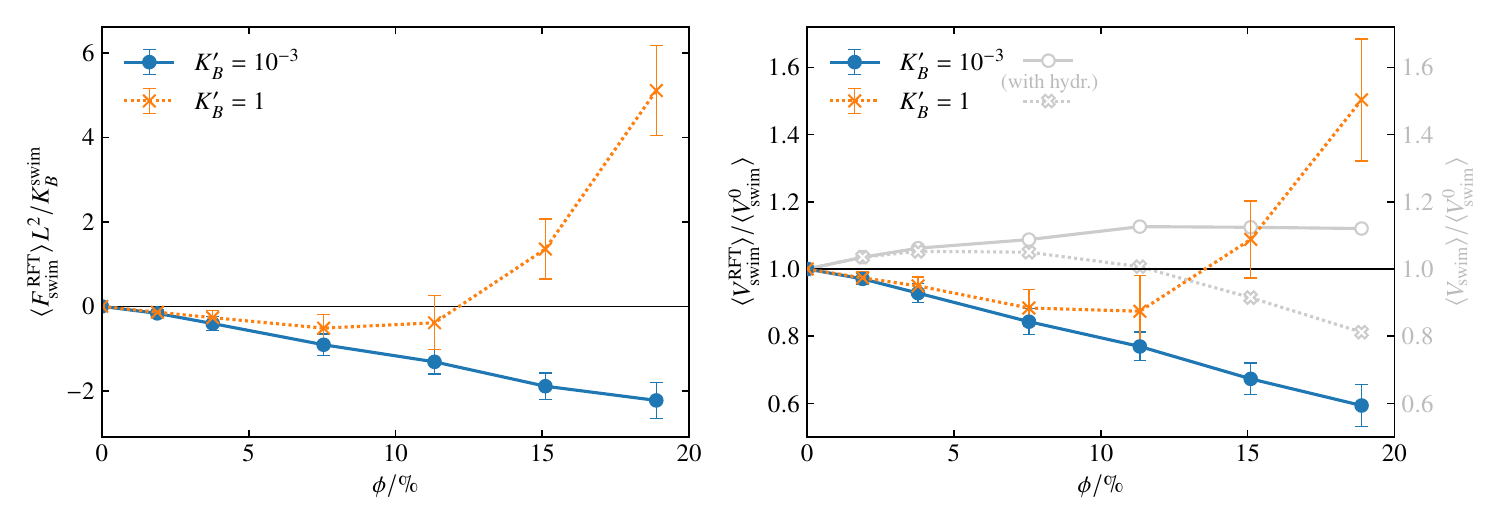}};
        \node[anchor=north west,inner sep=0] at (0,-0.2) {(\textit{a})};
        \node[anchor=north west,inner sep=0] at (8.3,-0.2) {(\textit{b})};
    \end{tikzpicture}    
    \caption{Non-hydrodynamic simulations using resistive force theory (RFT). In the absence of hydrodynamic interactions, the forces experienced by the swimmer strongly correlate to the swimmer speed. %
    (\textit{a}) Mean force on the swimmer in the swimming direction.
    (\textit{b}) Mean normalised swimmer speed (with results from \cref{fig:swimmer-speed-over-time}, which include hydrodynamic interactions, reproduced in the background in light grey for comparison).
     In both panels, results are again averaged from 20 periods of undulation and are plotted as a function of the volume concentration of network filaments, for both values of the network filament bending modulus. Results for each parameter set are averaged from 10 independent simulations, and error bars signify $\pm1$ standard deviation over these simulations.
    }
    \label{fig:mean-speed-over-phi-rft}
    \label{fig:forces-push-back-and-deform-rft}
\end{figure}

The resulting force and speed of the swimmer in these simulations without hydrodynamics are shown in  \cref{fig:mean-speed-over-phi-rft}, and, similar to the gait analysis simulations in \cref{fig:gait-sucks}, show no modest increase in speed at low-to-mid concentrations. Unlike simulations with full hydrodynamics (reproduced from \cref{fig:swimmer-speed-over-time} in light grey in the background of \cref{fig:mean-speed-over-phi-rft}), we see that the swimming speed in this concentration regime is strongly affected by the steric forces, as evidenced by the strong correlation between these forces (panel \textit{a}) and the resulting speed (panel \textit{b}).  At these concentrations, also observe the reduction in swimmer amplitude present in the full simulations, \cref{fig:amplitude-over-phi}.
    
Although the general shape of the total force graph is similar between RFT and the full simulations, see \cref{fig:mean-speed-over-phi-rft,fig:forces-push-back}, it is noticeable that the backwards force response of $K'_B=10^{-3}$ is stronger, and of $K'_B=1$ is weaker, without hydrodynamics.  This can be attributed to the difference in hydrodynamic model since when hydrodynamics are included, both the swimmer and the suspended filaments can deform each other without contact; whereas in the drag-based model, the filaments only affect each other on contact through the repulsive forces.  In short, the presence of hydrodynamics in \cref{fig:swimmer-speed-over-time} significantly changes the swimmer behaviour from a steric force driven regime seen in the absence of hydrodynamics, in \cref{fig:mean-speed-over-phi-rft}.

\subsection{Summary of investigation}
In this examination of swimmer motion, we have analysed the swimming gait and the forces experienced by the swimmer.  While we find that the fluctuations in the swimming speed are closely correlated with collisions between the swimmer and filaments, enhancement in the mean speed is not linked to the force experienced by the swimmer or changes in swimming gait.  In fact, both effects would produce a monotonic decrease in the speed with filament concentration.  This suggests that the enhancement is instead linked to hydrodynamic interactions between the swimmer and filaments.  We confirm this by performing resistive force theory simulations where hydrodynamic interactions between the swimmer and filaments are removed and for which we do not observe any enhanced speed.

\section{Discussion and conclusions}
\label{sec:discussion}

In this paper, we performed simulations examining the changes in swimming speed experienced by an undulatory swimmer in filament suspensions and immersed filament networks.  The simulations reveal a non-monotonic dependence of the speed on the filament concentration, with initial increases of the swimming speed to peak values of 5\%--15\% greater than the speed in a viscous fluid.  These increases in speed are in line with predictions of swimming speeds using continuum models to describe the surrounding medium.  Filament and connection flexibility are required to enhance the speed.  Filaments and connections that are too stiff will instead hinder locomotion.  Along with the speed, we examine speed fluctuations, which cannot be captured with continuum models.  These fluctuations are found to be significant and increase monotonically with filament concentration.  

To understand these changes in swimmer motion we examined closely the changes in swimming gait and force experienced by the swimmer.  Although contact between the swimmer and its environment correlate with speed fluctuations, we cannot attribute the mean speed increase to these interactions, or indeed to changes in the swimming gait, since we see that independently these effects lead to a monotonic decrease in speed with filament concentration. Instead, we attribute the speed enhancement to hydrodynamic interactions between the swimmer and the filaments in the environment by showing that when we remove them, in resistive force theory simulations, we no longer observe enhanced speed.

Previous work on planar undulatory motion between tethered obstacles showed a tenfold increase in the swimming speed in structured environments \citep{majmudar_experiments_2012}, and an increase of a factor of two- to three-times in unstructured media \citep{kamal_enhanced_2018}.  By contrast, our results here are significantly more modest.  We attribute this more modest swimming speed enhancement to the freedom of movement afforded by three dimensions, and to a difference in the fixing of the obstacles.  The previous work \citep{majmudar_experiments_2012,kamal_enhanced_2018} showed that planar motion was primarily governed by steric contact: motion could be hindered by trapping, and could be promoted by the provision of obstacles for the swimmer to push against. In contrast, three-dimensional motion allows for escape from constrained environments, diluting the ability for the structure to hinder or promote motion directly. In the earlier work, the obstacles are fixed to a point in space, allowing the forces on them to not be completely resolved by the fluid. Here, conversely, the forces on the suspended filaments are completely balanced by the surrounding fluid. Finally, we have concluded that speed enhancement in the current work is instead linked to hydrodynamic interactions with neighbouring filaments; an increased number of neighbours is another property of three-dimensional geometry. 

We also see less evidence of frequent or sharp turns in the swimmer trajectories when compared to planar motion, resulting in swimming paths that are more linear and less diffusive.  We note, however, that our simulations of $20T$ are well below the correlation time, $\tau_c$, which we estimate by fitting $\langle \uv{P}(s)\cdot\uv{P}(s+t) \rangle$ with $\exp(-t/\tau_c)$.  We find values of $\tau_c$ between $110T$ and $3500T$ in our parameter regime.  Nonetheless, we find trends which agree with the planar study: increasing the concentration and stiffness of network filaments results in greater changes in swimmer orientation, reflecting that the swimmer is deflected more by larger numbers of suspended filaments.

The motivation for our work was to model a complex fluid, such as mucus, as an immersed three-dimensional filament network immersed in as viscous fluid, and elucidate the effects of physical interactions with the network on swimmer motion, especially with regard to the conditions under which this model recovers behaviour seen in viscoelastic experiments and in continuum models.  Our results on swimming speeds, with mild speed enhancement at low concentrations of flexible networks but hindrance in stiff, concentrated networks, display the same trends as results obtained using two- and three-dimensional viscoelastic continuum models when the Deborah number or an analogous measure of resistance is increased. Interestingly, in these studies, either the waveform amplitude is relatively fixed \citep{teran_viscoelastic_2010,thomases_mechanisms_2014}, or can correlate faster speeds with increased (free to vary) amplitudes \citep{leiderman_swimming_2016,olson_effect_2015}; whereas the speed enhancement we observe happens despite a decrease in the swimming stroke amplitude.  Along with these studies based on continuum descriptions, our results are in line with those obtained examining undulatory swimming through highly connected spring networks \cite{wrobel_enhanced_2016}. This study, where the swimmer waveform is prescribed and the network interact with the fluid only through the nodes, speed increases of approximately 20\% were also reported.  We see a similar speed increases, especially in the more flexible networks, where the swimmer waveform is mainly unaffected by the suspended filaments, as in \cite{wrobel_enhanced_2016}.

The density and spacing of the network filaments in the simulations presented here were inspired by that found in mucin networks.  Although we have considered a large range of concentrations, typical mucin concentration in mucus is between 0.2\% and 5.0\% v/w; water forms approximately 95\% \citep{leal_physicochemical_2017,gniewek_coarse-grained_2010}.  And in this regime, we have found mild speed promotion for an undulatory swimmer in both flexible and stiff filament environments, with a small further improvement with stiffer springs in the networks.
This is consistent with positive sperm migration in human cervical mucus in the ovulatory phase of the menstrual cycle, when hydration increases to 98\% w/w and the mucin fibres can appear globular and disconnected \citep{curlin_cervical_2013,morales_human_1993,bergman_spermigration_1953,brunelli_globular_2007}. The luteal phase, correlating with highly reduced migration, sees hydration decreasing to 90\% w/w and the fibres appearing densely connected \citep{katz_analysis_1997,brunelli_globular_2007}. In this less dilute regime, we begin to see speed reduction in stiff filament networks, which is enhanced by stiffer network connecting springs.  It is important to note that while the chemical composition of mucins and polymers are well-known \citep{leal_physicochemical_2017}, their mechanical properties at an individual fibre level, as well as the mechanical properties of their connections, are less well understood \citep{lai_altering_2009}.

Classes of microswimmers are able to propel themselves through different locomotive styles: here we have found that low-to-mid concentrations of flexible networks are mildly favourable towards the undulatory motion of microorganisms such as sperm cells or \emph{C.\ elegans}. 
Another significant locomotive style is to rotate helical flagella, seen in \emph{P.\ aeruginosa} or \emph{E.\ coli}. Simulations of this motion, resolving the fluid using alternative methods such as multiparticle collision dynamics or regularised Stokeslets have seen speed promotion up to 60\% in confined tubes \citep{lagrone_elastohydrodynamics_2019} and polymer solutions \citep{zottl_enhanced_2019}. In the latter case, speed up is attributed to chirality of the flagellum, which is absent in undulatory swimmers. This is interesting because one purpose of biological networks of the kind modelled here, such as mucus, is to block pathogens. We might expect that these networks are favourable to certain locomotive styles over others, and it remains as a future direction to determine this within the framework presented here.

\section*{Acknowledgements}
The authors gratefully acknowledge support from EPSRC Grant EP/P013651/1.

\appendix
\section{Rigid body motion}
\label{sec:rbm}

In order to assess the effectiveness of the swimming gait that emerges in the filament environment, we determine the swimming speed produced by the gait in a viscous fluid absent of any filaments.

\subsection{Extracting the swimming gait}

Suppose that in the filament suspension, the swimmer's centre of mass has velocity $\v{u}^{\text{sus}} = \fd{ \v{Y}_{\text{CoM}}}{t}$ and angular velocity $\v{\omega}^{\text{sus}}$. Note that we use lowercase letters to represent the rigid body motion (vectors of length 3 in 3D) and uppercase letters for the segments of the swimming filament (indexed by $n$); the notation `$\text{sus}$' indicates this data is from the suspension simulation.  The motion of segment $n$ can then be written as
\begin{align}
    \v{V}_n^\text{sus} &= \tilde{\v{V}}_n + \v{u}^{\text{sus}} + \v{\omega}^{\text{sus}}\times\Delta\v{Y}_n, \label{v-to-sum}\\
    \v{\Omega}_n^\text{sus} &= \tilde{\v{\Omega}}_n + \v{\omega}^{\text{sus}}, \label{om-to-rearrange}
\end{align}
where $\Delta\v{Y}_n = \v{Y}_n - \v{Y}_{\text{CoM}}$; and $\tilde{\v{V}}_n$ and $\tilde{\v{\Omega}}_n$ are the segment velocities and angular velocities associated with swimmer deformation for swimming.  

To extract $\tilde{\v{V}}_n$ and $\tilde{\v{\Omega}}_n$, we need to find expressions for $\v{u}^\text{sus}$ and $\v{\omega}^\text{sus}$ in terms of the known $\v{V}_n^\text{sus}$ and $\v{\Omega}_n^\text{sus}$.  Summing \cref{v-to-sum} over $n$, we find
\begin{equation}
    \v{u}^\text{sus} = \frac{1}{N}\sum_n \v{V}_n^\text{sus}, 
\end{equation}
since we must have $\sum_n\tilde{\v{V}}_n = \vu{0}$ and $\sum_n\Delta\v{Y}_n = \vu{0}$.

Now we consider taking the cross product of $\Delta\v{Y}_n$ and \cref{v-to-sum}, and summing:
\begin{align}
        \sum_n \Delta\v{Y}_n \times \v{V}_n^\text{sus} 
        &= 
        \sum_n \Delta\v{Y}_n \times \tilde{\v{V}}_n + 
        \sum_n \Delta\v{Y}_n \times \v{u}^{\text{sus}} +
        \sum_n \Delta\v{Y}_n \times (\v{\omega}^{\text{sus}}\times\Delta\v{Y}_n)\\
        &= \vu{0} + \vu{0} + \t{I}_M\v{\omega}^\text{sus},
\end{align}
where, on the right hand side, 
\begin{itemize}
    \item term 1 is zero by the definition of $\tilde{\v{V}}$, 
    \item term 2 is zero since $\sum_n\Delta{\v{Y}}_n = \vu{0}$,
    \item term 3 introduced the moment of inertia matrix, defined in terms of a skew-symmetric matrix $[\v{a}]_\times$ such that $[\v{a}]_\times\v{b}=\v{a}\times\v{b}$,
    \begin{equation}
        \t{I}_M = -\sum_n [\Delta\v{Y}_n]_\times[\Delta\v{Y}_n]_\times, \qquad [\v{a}]_\times \coloneqq \begin{pmatrix}0 & -a_3 & a_2 \\ a_3 & 0 & -a_1 \\-a_2&a_1&0\end{pmatrix}.
        \label{cross-product-matrix}
    \end{equation}
\end{itemize}
Hence,
\begin{equation}
    \v{\omega}^\text{sus} = \t{I}_M^{-1}\sum_n\Delta\v{Y}_n\times\v{V}_n^\text{sus}.
\end{equation}

We can therefore now rearrange \cref{v-to-sum,om-to-rearrange} to find an expression for the local motion, $\tilde{\v{V}}_n$ and $\tilde{\v{\Omega}}_n$, as a sum of fully evaluated terms:
\begin{subequations}
\label{v-and-om-tilde}
\begin{align}
    \tilde{\v{V}}_n &= \v{V}_n^\text{sus} - \v{u}^{\text{sus}} - \v{\omega}^{\text{sus}}\times\Delta\v{Y}_n,\\
    \tilde{\v{\Omega}}_n &= \v{\Omega}_n^\text{sus} - \v{\omega}^{\text{sus}}. 
\end{align}
\end{subequations}

We can now take these expressions as representing the gait of the swimmer, and place them into an empty fluid.

\subsection{Resulting swimming speed in a viscous fluid}

Now consider the swimmer in a viscous fluid free from any suspended filaments. The mobility problem is, as in \cref{eq:mobility-problem},
\begin{equation}
    \mathcal{M}
    \begin{pmatrix}
        \v{F} \\ 
        \v{T}
    \end{pmatrix} =
    \begin{pmatrix}
        \v{V} \\ 
        \v{\Omega}
    \end{pmatrix},
\end{equation}
where we have dropped the $H$ for `hydrodynamic' for notational convenience.
Recall that $\mathcal{M}$, for a single swimming filament formed of $N$ segments, has size $6N \times 6N$ in 3D. Once again we can decompose the motion into a rigid body motion and a local deformation for each segment $n$,
\begin{subequations}
\label{decompose-empty}
\begin{align}
    \v{V}_n
    &= \tilde{\v{V}}_n + \v{u} + \v{\omega}\times\Delta\v{Y}_n, \\
    \v{\Omega}_n&= 
        \tilde{\v{\Omega}}_n + \v{\omega},
\end{align}
\end{subequations}
where $\tilde{\v{V}}_n$ and $\tilde{\v{\Omega}}_n$ are specified from \cref{v-and-om-tilde} and the unknowns are $\v{F}_n$, $\v{T}_n$, $\v{u}$ and $\v{\omega}$. We solve this system for the unknowns, subject to the conditions that the total force and torque are zero,
\begin{align}
     \sum_n\v{F}_n & = \vu{0},\label{empty-f0}\\
    \sum_n(\Delta\v{Y}_n \times \v{F}_n + \v{T}_n) &= \vu{0}\label{empty-t0}.
\end{align}

\Cref{decompose-empty,empty-f0,empty-t0} can be combined to form a system of equations with a symmetric matrix of size $(6N+6)\times(6N+6)$,
\begin{equation}
    \begin{pmatrix}
     &  & -\t{I} & [\Delta \v{Y}]_\times\\
    \multicolumn{2}{c}{\smash{\raisebox{0.5\normalbaselineskip}{$\mathcal{M}$}}} & \tu{0} & -\t{I}\\
    -\t{I} & \tu{0} & \tu{0} & \tu{0} \\
    [\Delta \v{Y}]_\times^\trans & -\t{I} & \tu{0} & \tu{0}
    \end{pmatrix}
    \begin{pmatrix}
    \v{F} \\
    \v{T} \\
    \v{u} \\
    \v{\omega}
    \end{pmatrix}
    = 
    \begin{pmatrix}
    \tilde{\v{V}} \\
    \tilde{\v{\Omega}} \\
    \vu{0} \\
    \vu{0} 
    \end{pmatrix},
\end{equation}
recalling that $\tilde{\v{V}}^\trans = (\tilde{\v{V}}_1^\trans,\dots,\tilde{\v{V}}_N^\trans)$ and similarly for $\tilde{\v{\Omega}}$, $\v{F}$ and $\v{T}$. Here, $\t{I}$ is the identity matrix and $[\Delta \v{Y}]_\times$ is again the skew-symmetric matrix corresponding to the cross product from \cref{cross-product-matrix}, in both cases suitably stacked for each segment $n$.  Solving this system gives the swimming velocity in the viscous fluid.

\section{Resistive force theory}
\label{sec:rft}

Resistive force theory (RFT) \citep{lauga_hydrodynamics_2009, johnson_flagellar_1979} can be used to provide the mobility matrix in \cref{eq:mobility-problem}.  In RFT, the filament segments are hydrodynamically uncoupled, with the force and torque on a segment dependent only on the segment's orientation, velocity and angular velocity.
In this way, it can be thought of as a drag-based model, but one where the force--velocity drag coefficients are different in the directions normal and tangent to the segment orientation. 

\begin{figure}
    \centering
	\includegraphics[width=\textwidth]{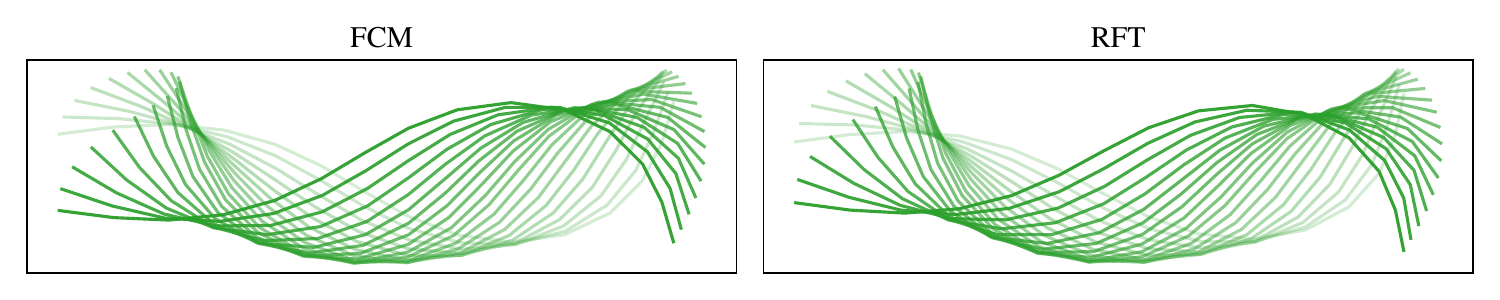}
	\caption{Comparison of the waveform of a single swimmer in an empty fluid with full hydrodynamics (FCM), and with the drag-based resistive force theory (RFT). The swimming direction is to the right and the axes are the same for both figures. Half an undulation period is depicted, and the centreline fades as the snapshot retreats into the past.}
    \label{fig:swimmer_snapshots_overlay_rft}
\end{figure}

In the mobility formulation of RFT, the velocity and angular velocity of segment $n$ are given by 
\begin{align}
    \v{V}_n &= \left[ \alpha_{||}\uv{t}_n\uv{t}_n^{\,\trans} + \alpha_\perp (\t{I}- \uv{t}_n\uv{t}_n^{\,\trans}) \right] \v{F}_n,\\
    \v{\Omega}_n &= \beta \v{T}_n,
\end{align}
where $\alpha_{||}$, $\alpha_{\perp}$, $\beta$ are mobility coefficients for the segments. The torque coefficient $\beta$ is simply that for a sphere in Stokes flow,
\begin{equation}
    \beta = (8\upi a^3 \mu)^{-1},
\end{equation}
while $\alpha_{||}$ and $\alpha_{\perp}$ are initially estimated using FCM simulations of straight filaments, following the approach of \citet[supplementary material]{schoeller_flagellar_2018}. These two parameters are then tuned so that the swimming speed and maximum beating amplitude match the full FCM simulation for a single swimmer in an empty fluid. Doing so, we choose the numerical values
\begin{equation}
    \alpha_{||} = 0.1177(a\mu)^{-1}, \quad \alpha_{\perp} = 0.08208(a\mu)^{-1}.
\end{equation}
A comparison of the resultant waveform for FCM and RFT is given in \cref{fig:swimmer_snapshots_overlay_rft}. The ratio $\alpha_{||}/\alpha_{\perp} \approx 1.44$ is similar to previous work \citep{schoeller_flagellar_2018}. While for an infinitely thin filament, we expect a ratio of 2 \citep{johnson_flagellar_1979}, we observe that as the swimmer has an aspect ratio of $2a/L \approx 0.06$, we are not in this limiting regime. Note that RFT can also be presented in the resistance formulation, where this ratio is presented in terms of drag coefficients, $\xi_\perp/\xi_{||}$.

\bibliographystyle{jfm2}
\bibliography{references.bib}

\end{document}